%% file: p.tex
\documentclass[format=acmsmall, anonymous=false, screen=true]{acmart}
\renewcommand\footnotetextcopyrightpermission[1]{}
\input{macros}

\input{packages}

\setlist[itemize,1]{leftmargin=5mm,itemsep=0mm}
\setlist[enumerate,1]{leftmargin=5mm,itemsep=0mm}

\definecolor{keywords}{HTML}{dd1c77}
\definecolor{strings}{HTML}{00999A}
\definecolor{comments}{HTML}{31a354}
\definecolor{identifiers}{HTML}{000000}
\begin{document}


\title{Rank/Select Queries over Mutable Bitmaps}
\author{Giulio Ermanno Pibiri}
\affiliation{%
 \institution{ISTI-CNR, Pisa, Italy}
}
\email{giulio.ermanno.pibiri@isti.cnr.it}
\author{Shunsuke Kanda}
\affiliation{%
 \institution{RIKEN AIP, Tokyo, Japan}
}
\email{shunsuke.kanda@riken.jp}
\input{abstract}
\maketitle
\renewcommand{\shortauthors}{G.\,E. Pibiri and S. Kanda}

\input{introduction}
\input{related}

\input{setup}

\input{overview}

\input{prefix_sums}

\input{small_bitmaps}
\input{final_result}

\input{conclusions}

\section{Acknowledgments}
The first author was partially supported by the BigDataGrapes (EU H2020 RIA, grant agreement N\textsuperscript{\b{o}}780751) and the OK-INSAID (MIUR-PON 2018, grant agreement N\textsuperscript{\b{o}}ARS01\_00917) projects.

\balance
\renewcommand{\bibsep}{3.0pt}
\bibliographystyle{ACM-Reference-Format}
\bibliography{bibliography}

\end{document}

%% file: macros.tex
\newcommand{\parag}[1]{\vspace{0.2cm}\noindent\textbf{{#1.}}}

\newcommand{\code}[1]{\mbox{\texttt{#1}}}
\newcommand{\var}[1]{\mbox{\texttt{#1}}}
\newcommand{\bit}[1]{\mbox{\texttt{#1}}}

\usepackage{xcolor, soul}

\usepackage{pbox}
\usepackage{makecell}
\usepackage{tikz}
\usetikzlibrary{calc}

\newcommand{\method}[1]{\textup{\textsf{#1}}}
\newcommand{\Sum}{\code{sum}}
\newcommand{\Update}{\code{update}}

\newcommand{\Search}{\code{search}}
\newcommand{\Rank}{\code{rank}}
\newcommand{\Select}{\code{select}}
\newcommand{\Flip}{\code{flip}}
\newcommand{\Xor}{\textsf{XOR}}

\newcommand{\fentree}{\method{Fenwick-Tree}}
\newcommand{\segtree}{\method{Segment-Tree}}

\newcommand{\ceil}[1]{\lceil{#1}\rceil}
\newcommand{\floor}[1]{\lfloor{#1}\rfloor}

\newcommand{\SelectW}{\code{select64}}

%% file: packages.tex
\usepackage{graphicx}
\usepackage{balance}
\usepackage{xcolor}
\usepackage{url}
\usepackage{amsmath}
\usepackage{multirow}
\usepackage{booktabs}
\usepackage{makecell}
\usepackage[caption=false]{subfig}
\usepackage{wrapfig}
\usepackage{xcolor, soul}
\usepackage{siunitx}
\usepackage{enumitem}
\usepackage{rotating}
\usepackage{tablefootnote}
\usepackage[noend]{algpseudocode}
\usepackage[linesnumbered
			,vlined]{algorithm2e}
\usepackage{listings}
\usepackage{afterpage}

%% file: abstract.tex
\begin{abstract}
The problem of answering \emph{rank/select} queries over a bitmap
is of utmost importance for many succinct data structures.
When the bitmap does not change, many solutions exist in the
theoretical and practical side.
In this work we consider the case where
one is allowed to \emph{modify} the bitmap via a
$\Flip(i)$ operation that toggles its $i$-th bit.
By adapting and properly extending some results concerning
\emph{prefix-sum} data structures,
we present a practical solution to the problem, tailored for
modern CPU instruction sets.
Compared to the state-of-the-art, our solution improves runtime
with no space degradation.
Moreover, it does not incur in a significant runtime penalty
when compared to the fastest immutable indexes,
while providing even lower space overhead.
\end{abstract}

%% file: introduction.tex
\section{Introduction}\label{sec:introduction}

Given a bitmap $\mathcal{B}$ of $u$ bits where $n$ are set,
the query $\Rank(i)$ asks for the number of bits set in $\mathcal{B}[0..i]$,
for an index $0 \leq i < u$;
the query $\Select(i)$ asks for
the position of the $i$-th bit set,
for $0 \leq i < n$
(in our formulation and throughout the paper, all indexes start from 0
so that if $\Select(i)=k$, then $\Rank(k)=i+1$).
These queries are at the core of more complicated
\emph{succinct} data structures~\cite{jacobson1988succinct},
whose objective is to support
efficient operations on theoretically-optimal compressed data, i.e.,
within lower order terms from the theoretic minimum space.
Important applications of succinct data structures
include string dictionaries~\cite{martinez2016practical,kanda2017string},
orthogonal range querying~\cite{brisaboa2013space} and filtering~\cite{zhang2018surf},
Web graphs representation~\cite{brisaboa2014compact},
text indexes~\cite{ferragina2009compressed},
inverted indexes~\cite{pibiri2020techniques},
just to name a few.
For this reason, many solutions were proposed
in both theory and practice
to answer these queries efficiently with small extra space.
It is intuitive that a data structure built on top of $\mathcal{B}$ --
hereafter, named the \emph{index} --
helps in answering both queries faster, thus indicating that many
space/time trade-offs are possible.

On the theoretical side, each of the two queries can be answered in $O(1)$ time
with a \emph{separate} index that just takes $o(u)$ extra bits~\cite{Clark96,jacobson1988succinct}.
When \emph{both} queries are to be supported using the \emph{same}
$o(u)$-bit index instead,
it is not possible to do better than $O(\log u / \log w)$, where $w$ is the
machine word size in bits~\cite{FS89}.
The $o(u)$ term in space occupancy is also almost optimal~\cite{miltersen2005lower,golynski2007optimal}.
Despite their elegance, these results are of little usefulness in practice
because their corresponding indexes are either too complicated
(i.e., with high constant factors and unpredictable memory accesses
that impair efficiency) or take much space.
Instead, practical solutions have now reached a mature state-of-the-art:
depending on the size of the bitmap and the design of the index,
both queries can be answered in (roughly speaking) tens of nanoseconds
with $3-28\%$ of extra space.

Known solutions assume that the bitmap $\mathcal{B}$ is immutable.
In this work we consider an extension to the problem defined above,
where one
is allowed to modify $\mathcal{B}$ with a $\Flip(i)$ operation
that toggles $\mathcal{B}[i]$, the $i$-th bit of $\mathcal{B}$:
the bit is set to \bit{1} if it was \bit{0} and vice versa,
i.e., $\mathcal{B}[i] = \Xor(\mathcal{B}[i],\bit{1})$.
We call this problem:
\emph{rank/select queries over mutable bitmaps}.
We leave the bitmap uncompressed and
we do not take into account compressed representations.

For example, suppose $\mathcal{B}$ is \bit{[01101101010101110]},
with $u=17$ and $n=10$.
Then, $\Rank(7) = 5$ and $\Select(7) = 13$. If we perform $\Flip(3)$ and
$\Flip(6)$, then $\mathcal{B}$ becomes
\bit{[011\textbf{1}11\textbf{1}1010101110]},
so that now $\Rank(7) = 7$ and $\Select(7) = 9$.

Clearly, the index data structure should reflect the changes
made on $\mathcal{B}$ and this complicates the problem compared
to the immutable setting.
We cannot hope of doing better than, again,
$O(\log u / \log w)$ time for this problem,
due to a lower bound by~\citet*{FS89}.
However, it is not clear
what can be achieved in practice, which is our concern with this work.
The problem is relevant for succinct dynamic data structures~\cite{prezza2017framework},
and other applications such as
counting transpositions in an array and generating graphs by preferential
attachment~\cite{marchini2020compact}.

\parag{Our Contribution}
Our solution to the problem builds on two main ingredients:
(1) a space-efficient bitmap index that
leverages SIMD instructions~\cite{SIMDIntel} and other optimizations
for fast query evaluation, and
(2) efficient algorithms to solve the problem
of rank/select over small, immutable, bitmaps.
Regarding point (1),
we adapt and extend a recent result~\cite{pibiri2020practical}
about the $b$-ary {\segtree},
and show that SIMD is very effective to search and update the data structure.
We describe two variants of the data structure, tailored
for both the widespread AVX2 instruction set and the new AVX-512 one.
As per point (2), we review, extend, and compare all existing approaches
to perform rank/select over small bitmaps (e.g., 256 and 512 bits).
These approaches include
broadword techniques, intrinsics, SIMD instructions,
or a combination of them.
The results of the comparison may be of independent interest
and relevant to other problems.

The best results for the two previous points are then combined to tune
a mutable bitmap data structure with support for rank/select queries,
with the aim of establishing new reference trade-offs.
Compared to the state-of-the-art~\cite{marchini2020compact}
(available as part of the \textsf{Sux} library~\cite{sux}),
our solution is faster and consumes a negligible amount of
extra space.
We also compare the data structure against several other solutions
for the immutable setting,
implemented in popular libraries  such
as \textsf{Sdsl}~\cite{gog2014theory,gog2014optimized},
\textsf{Succinct}~\cite{grossi2013design},
and \textsf{Poppy}~\cite{zhou2013space}.
Our data structure is competitive with the fastest indexes
and takes less space,
while allowing clients to alter the bitmap as needed.

We publicly release the C++ implementation of the solutions proposed
in this article at
\url{https://github.com/jermp/mutable_rank_select}.

%% file: related.tex
\section{Related Work}\label{sec:related_work}

In this section we review solutions devised to solve
the problem of ranking and selection over mutable
and immutable bitmaps.

\subsection{Mutable Bitmaps}

To the best of our knowledge, only~\citet*{marchini2020compact}
addressed the problem of rank/select over mutable
bitmaps, providing a solution that uses
a (compressed) Fenwick tree~\cite{fenwick1994new} as index.
Therefore, this will be our direct competitor for
this work.
(To be fair, also the \textsf{Dynamic} library by~\citet*{prezza2017framework}
contains a solution based on a B-tree but it is several times
slower than the solution by~\citet*{marchini2020compact},
so we excluded it from our experimental analysis.)

\begin{table}[t]
\centering
\caption{
Summary of data structures.
A space overhead indicated with $x-y\%$ is achieved by varying
the density of the bitmap from 0.1 to 0.9.
}
\subfloat[{\Rank}]{
\scalebox{1.0}{\input{tables/summary_rank}}
\label{tab:summary_rank}
}\\
\subfloat[{\Select}]{
\scalebox{1.0}{\input{tables/summary_select}}
\label{tab:summary_select}
}
\label{tab:summary}
\end{table}

\subsection{Immutable Bitmaps}\label{sec:immutable_bitmaps}

Solutions devised for the problem in the immutable setting abound
in the literature. We summarize the main different approached
in Table~\ref{tab:summary}.

\parag{Rank}
The first data structure for constant-time {\Rank} was introduced by~\citet*{jacobson1988succinct}, and consists of a two-level index
taking $O(u \log\log u / \log u) = o(u)$ bits of extra space.
The index stores precomputed rank values for blocks of the bitmap,
so that a query is resolved by accessing a proper value in each
of the two levels.
\citet*{vigna2008broadword} provided a seminal implementation
of this data structure  for blocks of 512 bits --
known as \textsf{Rank9} --
by interleaving the data of the two levels to improve locality of reference.
The data structure consumes 25\% extra space.
\citet*{gog2014optimized} presented two variants of \textsf{Rank9}:
one consumes less space by resorting on larger block sizes;
($\approx$6\% extra space);
the other is more cache-friendly and interleaves the index with the
bitmap itself.
We refer to the former as \textsf{Rank9.v2} and to the latter as \textsf{IL}.

\citet{zhou2013space} developed \textsf{Poppy}, a similar data structure to
the space-efficient variant of \textsf{Rank9} by \citet*{gog2014optimized}.
The main difference is that \textsf{Poppy} uses an additional level for the index,
therefore allowing to design the two-level index using 32-bit integers,
instead of 64-bit values.
The extra space of \textsf{Poppy} is just 3\%.


\parag{Select}
The solutions developed for {\Select} can be broadly classified into two categories: \emph{rank-based} and \emph{position-based}.
A rank-based solution answers queries by binary-searching
an index developed for {\Rank} queries.
The advantage is that the \emph{same} index can be reused for both queries,
so that {\Select} can be performed with no (or small) additional space
to that of {\Rank}.
Instead, position-based solutions are able to only answer {\Select} queries,
thus, these take significant more space than rank-based approaches.

\citet{vigna2008broadword} accelerated the binary search
in the \textsf{Rank9} index
with an additional level that can quickly find the area to be searched,
hence performing a \emph{hinted} binary search.
The original implementation requires 12\% space overhead in addition
to that of \textsf{Rank9}, that is, a total space overhead of 37\%.
\citet{grossi2013design} also implemented the hinted binary search
with only 3\% extra space, thus, the total space overhead becomes 28\%.

A position-based solution builds an index storing sampled {\Select} results.
The first data structure for constant-time queries was introduced by \citet*{Clark96}
and consists of a three-level index taking $3u / \ceil{\log\log u} + O(\sqrt{u} \log u \log\log u)$ bits of space in addition to the bitmap.
However, a straightforward implementation consumes a lot of space, i.e., 60\% \cite{gonzalez2005practical}.
\citet*{gog2014optimized} simplified the implementation to reduce
the space overhead (which can be up to 20\%), and called it \textsf{MCL}.
\citet{okanohara2007practical} proposed the \textsf{DArray} data structure,
that essentially is Clark's solution with 2 levels instead of 3.

\citet{vigna2008broadword} proposed two data structures called
\textsf{Select9} and \textsf{Simple}.
The
\textsf{Select9} data structure adds selection capabilities on top of \textsf{Rank9}.
The space overhead depends on the \emph{density} of the bits set in the bitmap
(e.g., $25-32\%$ as reported by Vigna in his experimentation).
Although the space overhead is larger than 25\%, \textsf{Select9}
answers both queries.
Instead,
\textsf{Simple} does not depend on \textsf{Rank9} and builds a two-level index following a similar idea to that of the \textsf{DArray}.
The space overhead also depends on the density
(e.g., $9-46\%$).

\citet{zhou2013space} proposed \textsf{CS-Poppy} that adds selection
capabilities on top of \textsf{Poppy}, using the idea of
\emph{combined sampling}~\cite{navarro2012fast}.
Since the additional index requires only 0.4\% of extra space, the space overhead of \textsf{Poppy} is still very low,
while also supporting {\Select} queries.

\parag{Compressed Layouts}\label{par:compressed_bitmaps}
Although in this work we do not aim at compressing the bitmap $\mathcal{B}$
nor its index,
it is important to remark that a separate line of research studied
the problem of Rank/Select queries over compressed bitmaps.

\citet{okanohara2007practical} proposed the first practical data structure that
approaches $u H_0(\mathcal{B}) + o(u)$ bits of space,
where $H_0(\mathcal{B})\leq 1$ is the zero-th order empirical entropy of $\mathcal{B}$.
However, the $o(u)$ term can be very large, except for very sparse bitmaps with
densities below 5\%.
\citet{navarro2012fast} proposed a practical implementation of the
compression scheme by \citet{raman2007succinct}.
The data structure has a space overhead of around 10\% of $u$, on top of the entropy-compressed bitmap.

Of particular intrest for this work are the compressed solutions proposed by
\citet*{grabowski2018rank}, whose main concern is fast query support.
We briefly sketch their data structures.
In general, the bitmap is divided into blocks, and blocks are grouped
into superblocks.
For each block, a fixed-size header stores rank or select results.
Depending on how headers and blocks are compressed, different layouts
are derived and named:
basic (no compression);
basic with compressed header or \textsf{bch}
(headers written differentially with respect to the first one);
mono-pair elimination or \textsf{mpe} (a ``mono-pair'' is a block
of $\mathcal{B}$ consisting in
all \bit{1}s or \bit{0}s, thus not stored);
cache-friendly or \textsf{cf} (results and offsets interleaved in a
contiguous memory area).
When the indexes are built for {\Select}, two parameters are used:
a sampling value $\ell$ so that the result of $\Select(i\ell)$ is stored
directly and a threshold value $T$ which determines the sparseness
of the blocks.
A block is considered to be sparse when $\Select((i+1)\ell)-\Select(i\ell)>T$,
or dense otherwise.
For a sparse block, all the results of $\Select(j)$ for $i\ell < j < (i+1)\ell$
are stored directly;
for a dense block, the range of bits
$\mathcal{B}[\Select(i\ell)+1..\Select((i+1)\ell)-1]$ is stored.

Experiments using real-world datasets show that the query
time of their compressed layouts is competitive with that of the
uncompressed counterparts.

%% file: tables/summary_rank.tex
\begin{tabular}{cccc}
\toprule
Method & Index Design & Space Overhead & Reference \\
\midrule

\textsf{Rank9} & 2-level & 25\% & \cite[Sect. 3]{vigna2008broadword} \\
\textsf{Rank9.v2} & 2-level & 6.3\% & \cite[Sect. 3]{gog2014optimized} \\
\textsf{IL} & 1-level & 12.5\% & \cite[Sect. 3]{gog2014optimized} \\
\textsf{Poppy} & 3-level & 3\% & \cite[Sect. 3.1]{zhou2013space} \\


\bottomrule
\end{tabular}

%% file: tables/summary_select.tex
\begin{tabular}{cccc}
\toprule
Method & Index Design & Space Overhead & Reference \\
\midrule

\textsf{Rank9+Hinted} & 3-level & 28\% & \cite[Sect. 3.3]{grossi2013design} \\

Clark's & 3-level & 60\% & \cite{Clark96} \\

\textsf{MCL} & 2-level & $3-20\%$ & \cite[Sect. 4]{gog2014optimized} \\

\textsf{DArray} & 2-level & $6-51\%$ & \cite{okanohara2007practical} \\

\textsf{CS-Poppy} & 4-level & 3.4\% & \cite[Sect 3.2]{zhou2013space} \\


\bottomrule
\end{tabular}

%% file: setup.tex
\section{Experimental Setup}\label{sec:setup}


Throughout the paper we discuss experimental results,
so in this section we report the details of our setup and methodology.
All the experiments were run
on a single core of an Intel i9-9940X processor, clocked at 3.30 GHz.
The processor has two private levels of cache memory per core:
$2 \times 32$ KiB $L_1$ cache
(32 KiB for instructions and 32 KiB for data); 1 MiB for $L_2$ cache.
The third level of cache, $L_3$, is shared among all cores and
spans $\approx$19 MiB.
All cache lines are 64-byte long.
The processor has proper support for
SSE, AVX, AVX2, and AVX-512, instruction sets.

Runtimes are reported in nanoseconds,
measured by averaging among \num{1000000} random queries.
(In the plots, the minimum and maximum runtimes
draw a ``pipe'', with the marked line inside representing
the average time.)
Prior to measurement, an untimed run of queries
is executed to warm-up the cache of the processor.

The whole codebase used for the experiments
is written in C++ and available at
\url{https://github.com/jermp/mutable_rank_select}.
The code was tested under different UNIX platforms and compilers.
For the experiments reported in the paper, the code was compiled with
\texttt{gcc} 9.2.1 under Ubuntu 19.10 (Linux kernel 5.3.0, 64 bits),
using the flags
\texttt{-std=c++17 -O3 -march=native}.

%% file: overview.tex
\section{Overview}\label{sec:overview}

\begin{figure}[!t]
\centering
\begin{lstlisting}
template <typename SearchablePrefixSums, typename BlockType>
struct mutable_bitmap {
  static const uint64_t block_size = BlockType::block_size;
  static const uint64_t words_in_block = block_size / 64;

  void build(uint64_t const* in, uint64_t num_64bit_words) {
    uint64_t num_blocks = ceil(static_cast<double>(num_64bit_words) / words_in_block);
    vector<uint16_t> counts;
    counts.reserve(num_blocks);
    bits.resize(words_in_block * num_blocks, 0);
    memcpy(bits.data(), in, num_64bit_words * sizeof(uint64_t));
    for (uint64_t i = 0; i != bits.size(); i += words_in_block) {
      uint16_t val = 0;
      for (uint64_t j = 0; j != words_in_block; ++j) {
        val += __builtin_popcountll(bits[i + j]);  // number of bits set in bits[i + j]
      }
      counts.push_back(val);
    }
    index.build(counts.data(), counts.size());
  }

  inline bool access(uint64_t i) const;  // returns the i-th bit

  void flip(uint64_t i) {
    bool bit = access(i);
    uint64_t word = i / 64, offset = i % 64, block = i / block_size;
    bits[word] ^= 1ULL << offset;
    index.update(block, bit);
  }

  uint64_t rank(uint64_t i) const {
    uint64_t block = i / block_size, offset = i % block_size;
    return (block ? index.sum(block - 1) : 0) +
           BlockType::rank(&bits[words_in_block * block], offset);
  }

  uint64_t select(uint64_t i) const {
    auto [block, sum] = index.search(i);
    uint64_t offset = i - sum;  // i - index.sum(block - 1);
    return block_size * block + BlockType::select(&bits[words_in_block * block], offset);
  }

  private:
  SearchablePrefixSums index;
  vector<uint64_t> bits;
};
\end{lstlisting}
\caption{The C++ implementation of a mutable bitmap with
{\Flip}/{\Rank}/{\Select} capabilities.
\label{code:mbitmap}}
\end{figure}

An elegant way to solve the problem of Rank/Select over a mutable bitmap
is to divide the bitmap into
blocks and use, as index, a \emph{searchable prefix-sum} data structure
that maintains the number of bits set into each block.
It follows that operations {\Rank}/{\Select} can be implemented by
first reading/searching the counters of the index and,
then, concluding the query locally inside a block.
The {\Flip} operation is instead supported by ``adjusting''
some counters of the index and toggling a bit of the bitmap.
Therefore, the problem actually reduces to two distinct
sub-problems:

\begin{enumerate}

\item The searchable prefix-sum problem, defined as follows.
Given an array $A$ of $m$ integers,
we are asked to support the following operations on $A$:
$\Sum(i)$ returns the quantity $\sum_{k=0}^i A[k]$,
$\Update(i,\Delta)$ sets $A[i]$ to $A[i]+\Delta$,
and $\Search(x)$ returns the smallest $0 \leq i < m$ such that $\Sum(i) > x$.
(As explained shortly, we are only interested in the special case
where $\Delta \in \{-1,+1\}$.)

\item Rank/Select over ``small'' bitmaps (i.e., the blocks)
without auxiliary space.

\end{enumerate}

We address these sub-problems in Section~\ref{sec:prefix_sums}
and~\ref{sec:small_bitmaps} respectively.
The best results for these sub-problems are then combined to tune
the final result in Section~\ref{sec:final_result}.

The example code in Figure~\ref{code:mbitmap} -- the \code{mutable\_bitmap}
class --
illustrates how a solution to the two aforementioned sub-problems
can be exploited to solve the problem we are considering.
In fact, note that
{\Rank} uses the query \code{index.sum()};
{\Select} uses the query \code{index.search()};
and {\Flip} uses the operation \code{index.update()}
with values of $\Delta$
restricted to be either $+1/-1$ (one more/less bit set inside a block).

\parag{Block Size}
Before proceeding further,
an important design decision has to be discussed --
how to choose the \emph{block size}.
(Throughout the paper, we will refer to this quantity as $B$.)
The block size determines space/time trade-offs:
handling many small blocks can excessively enlarge the space (and query time) of
the prefix-sum data structure; on the other hand, blocks that are too large
take more time to be processed with a consequent query slowdown.
For recent processors, the \emph{minimum} block size should always be 64 bits
given that both {\Rank} and {\Select} can be answered with a constant
number of instructions over 64 bits
(we provide further details about this in Section~\ref{sec:small_bitmaps}).
For a reasonable space/time trade-off,
the block size should be a \emph{small} multiple of the word size, e.g.,
256 or 512.
For example, using blocks of 256 bits
we expect a reduction in
space overhead of (roughly) $256/64 = 4\times$ compared to the case of 64-bit blocks.
For the rest of the paper, we will experiment with two block
sizes -- 256 and 512 bits.
What makes these two sizes particularly appealing for modern processors
is that (1) as long as no more than 512 bits are requested to be loaded,
at most 1 cache-miss is generated
because 512 bits is the (typical) cache-line size, and
(2) we can process 256 and 512 bits of memory in parallel using
the SIMD AVX2 and AVX-512 instruction sets.

Lastly, as we will better see in the following,
this choice of block size plays an important role for the
efficiency of the searchable prefix-sum data structure
because some of its counters
are sufficiently small as to permit packing several of these
in a SIMD register for parallel processing.


%% file: prefix_sums.tex
\section{The Searchable Prefix-Sum Data Structure}\label{sec:prefix_sums}

In this section we adapt and extend
the $b$-ary {\segtree} data structure with parallel updates
proposed by Pibiri and Venturini
to solve the prefix-sum problem~\cite{pibiri2020practical}.
They determined that this data structure performs better
than a binary {\segtree}~\cite{bentley1977solutions,bentley1980optimal}
and solutions based on the {\fentree}~\cite{fenwick1994new}.
Before presenting our data structure, we
recall some preliminary notions.

Given an array $A$ of $m$ integers, an internal node of a
binary {\segtree} stores the sum of the elements in $A[i,j]$,
its left child that of $A[i,\lfloor (i+j)/2 \rfloor]$, and
its right child that of $A[\lfloor (i+j)/2 \rfloor+1,j]$.
The $i$-th leaf of the tree (from left to right) just stores $A[i]$ and the root
the sum of the elements in $A[0,m-1]$.
It is easy to generalize the tree to become $b$-ary: a node holds
$b$ integers, the $i$-th being the sum of the elements of $A$ covered by
the sub-tree rooted in its $i$-th child.
Different trade-offs between the runtime of {\Sum} and {\Update}
can be obtained by changing the solution adopted to solve such
operations over the $b$ keys stored in a node.

Here is, in short, the solution by~\citet*{pibiri2020practical}
assuming $A$ to be an array of (possibly \emph{signed})
64-bit integers.
Every node of the tree is a two-level data structure,
holding an array of $b$ keys, say $\code{keys}[0..b)$.
This array is divided into segments of $\sqrt{b}$ keys each,
and each segment stored in prefix-sum.
A $\code{summary}[0..\sqrt{b})$ array stores the sums
of the elements in each segment, again in prefix-sum.
(Precisely, the $i$-th entry of \code{summary} is responsible for
the segment $i-1$, and $\code{summary}[0]=0$.)
It follows that {\Sum} queries are answered in constant time
with $\Sum(i) = \var{summary}[i / \sqrt{b}] + \var{keys}[i]$;
{\Update} operations are resolved by updating the \code{summary}
array and the specific segment comprising the $i$-th key.
Importantly, updates can be performed in parallel
exploiting the fact that four 64-bit keys can be packed
in a SIMD register of 256 bits.

\begin{figure}[!t]
\centering
\includegraphics[scale=1]{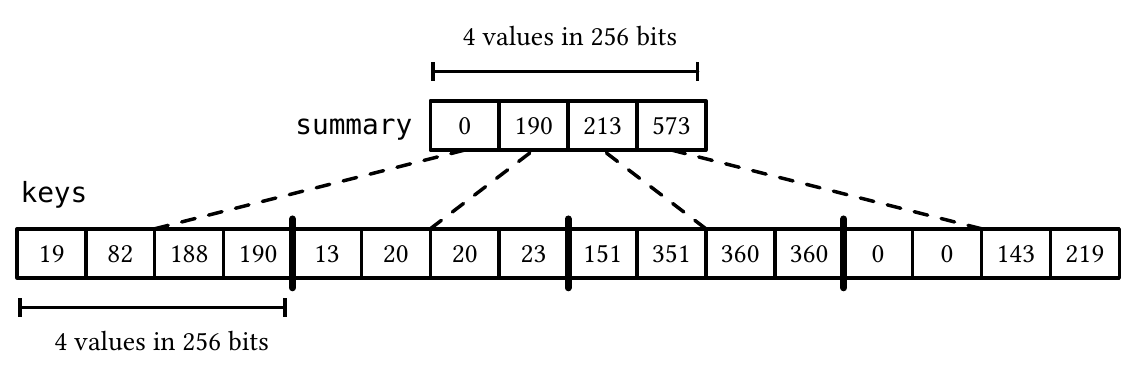}
\caption{A two-level node layout built from array
$[$19, 63, 106, 2, 13, 7, 0, 3, 151, 200, 9, 0, 0, 0, 143, 76$]$.
\label{fig:node16}}
\end{figure}

Figure~\ref{fig:node16} shows an example of this organization
for an input array of size $b=16$
containing only positive quantities as it is the case
for the problem we are interested in this work.

Once a node of the tree has been modeled in this way,
{\Sum} and {\Update} are resolved by traversing the tree
by issuing the corresponding operation on one node per each level.
Using reasonably large node fanouts $b$,
the resulting tree is very flat, e.g.,
of height $\lceil \log_{256} m \rceil$, where
$m$ is the array size and $b=256$.
Taking this into account,
\citet*{pibiri2020practical} also recommend to write a specialized code
path handling each possible value of tree height,
which avoids the use of loops and branches during
tree traversals for increased performance.
In what follows, we adopt their two-level node layout
and optimizations.

However, for the problem of rank/select over mutable bitmaps,
we need to solve a specialized version of the prefix-sum problem,
for the following reasons.
\begin{enumerate}
\item The input is an array of \emph{unsigned} integers,
with each value being at most 256 or 512
for our choice of block size.
\item The $\Delta$ value for {\Update} can only possibly be $+1$ or $-1$.
\item We also need the {\Search} operation, which we will implement using
SIMD instructions.
\end{enumerate}

Given these specifications,
we first describe our {\segtree} data structure
tailored for AVX2
and, then, discuss how to extend it to the new AVX-512 instruction set.
As explained before, we just need to care about the design
of the tree nodes since every tree operation is a sequence
of operations performed on the nodes.

\subsection{Using AVX2}
We start by highlighting some facts
that will guide and motivate
our design choices.

As per point (1) above, it follows that
the integers stored in the tree are all \emph{unsigned}.
This fact poses a challenge on the applicability of SIMD instructions
for updates and searches because most of such instructions
(up to and including the AVX2 set)
do not work with unsigned arithmetic.
For example, it is not possible to compare unsigned integers natively\footnote{The unsigned comparison can be simulated using a combination of instructions,
for a higher processing time.}
but only signed quantities.
Therefore, care should be put in ensuring that
the sign bit remains always 0 to guarantee correctness of the algorithms
which, in other words,
limits the dynamic of the integers that can be loaded into a SIMD register.
For example, we cannot ever store a quantity $v$ grater than $2^{15}-1$ in
a 16-bit integer, otherwise
the result of the comparison $v > x$ will be wrong
for any $0 \leq x < 2^{15}$.

To make the overall tree height as small as possible,
we would like to maximize the number
of integers that we can pack in a 256-bit SIMD register.
By construction of the {\segtree}, the deeper a node in the tree hierarchy,
the smaller the integers it holds, and vice versa.
This again limits the number of integers that can be processed at once
with a SIMD instruction at a given level of the tree.
Therefore we will define and use three node
types with \emph{different} fanouts: \code{node128}, \code{node64}, and \code{node32}.
Since their design is very similar,
in the following we give the details for one of them -- \code{node128} --
and summarize the different types in Table~\ref{tab:nodes}a.

\begin{figure}[!t]
\centering
\includegraphics[scale=0.6]{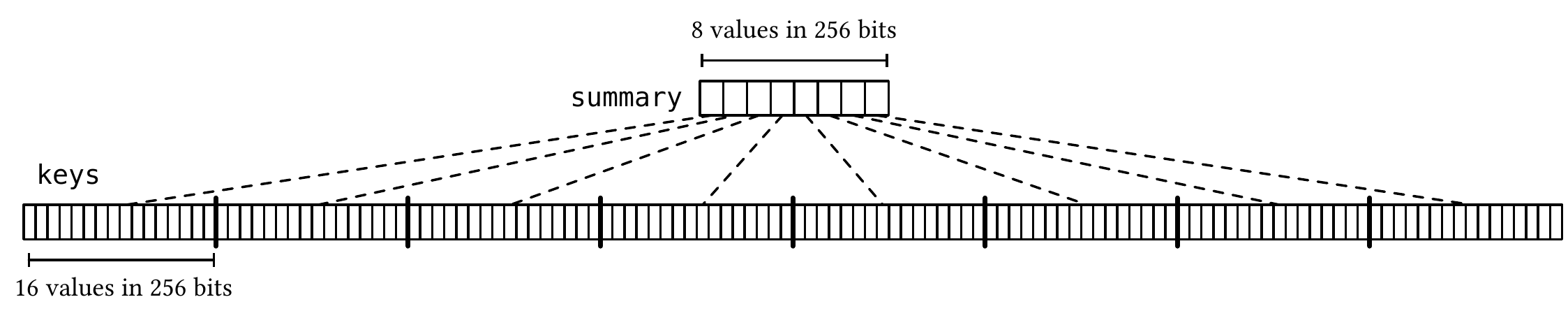}
\caption{The \code{node128} layout.
\label{fig:node128}}
\end{figure}

\parag{Node128}
For our choice of block size, each value of the input array can be
represented using 2 bytes.
Therefore we do so and
pack 16 values in a 256-bit node segment.
Let $B$ indicate the block size in bits.
Since each segment is stored in prefix-sum, it follows that
the maximum value we may store in a segment is $16B$.
For the discussion above, it follows that $B$ must satisfy
$16B < 2^{15}$ to guarantee correctness.
That is, every block size $B < 2^{11}$ bits does not violate
our design.
Then, we group 8 segments so that the maximum value stored in a segment
is at most $(8-1) \times 16B$ which we store into a 32-bit integer.
Summing up, we have 8 segments of 16 keys each, for a total
of 128 keys per node. We call this two-level node layout \code{node128}.
Figure~\ref{fig:node128} is a pictorial representation
of this layout.
We now illustrate how the methods, {\Sum}, {\Update}, and {\Search},
are supported with \code{node128}.

Queries are resolved as
$\Sum(i) = \var{summary}[i / 8] + \var{keys}[i]$.
The code implementing {\Update} is given below.
\begin{lstlisting}
void update(uint64_t i, bool sign) {
  if (i == fanout) return;
  uint64_t j = i / 16, k = i % 16;
  __m256i s1 = _mm256_load_si256((__m256i const*)T_summary + j + sign * 8);
  __m256i r1 = _mm256_add_epi32(_mm256_loadu_si256((__m256i const*)summary), s1);
  _mm256_storeu_si256((__m256i*)summary, r1);
  __m256i s2 = _mm256_load_si256((__m256i const*)T_segment + k + sign * 16);
  __m256i r2 = _mm256_add_epi16(_mm256_loadu_si256((__m256i const*)(keys + j * 16)), s2);
  _mm256_storeu_si256((__m256i*)(keys + j * 16), r2);
}
\end{lstlisting}
It uses two pre-computed tables.
Precisely, \code{T\_summary} stores $2 \times 8 \times 8$, 32-bit, signed integers,
where $\code{T\_summary}[i][0..7]$ is an array whose first $i+1$ integers
are 0 and the remaining $8 - (i \bmod 8) - 1$ are: $+1$ for $i=0..7$, and $-1$ for $i=8..15$.
The table \code{T\_segment}, instead, stores $2 \times 16 \times 16$, 16-bit, signed integers,
where $\code{T\_segment}[i][0..15]$ is an array where the first $i$ integers
are 0 and the remaining $16-(i \bmod 16)$ are: $+1$ for $i=0..15$, and $-1$ for $i=16..31$.
The tables are used to obtain the register
configuration that must be added to the summary and segment, respectively,
regardless the sign of the update.
(The difference with respect to the general approach
by~\citet*{pibiri2020practical} is that tables can be used to directly
obtain the register configuration to add because $\Delta$ is either $+1$ or $-1$
in this case, rather than a generic 64-bit value.)

Also note that the input parameter
\code{sign} is 0 for $\Delta=+1$ and 1 for $\Delta=-1$, so that
the use of {\Update} in the {\Flip} method of the \code{mutable\_bitmap} class
in Figure~\ref{code:mbitmap} (at page~\pageref{code:mbitmap}) is correct.


The $\Search(x)$ operation, that computes the smallest $i$ in $[0..128)$
such that $\Sum(i) > x$, can also be implemented using the SIMD instruction
\code{cmpgt} (compare greater-than).
Our approach is coded below, assuming that $x < \Sum(128-1)$.
\begin{lstlisting}
std::pair<uint64_t, uint64_t> search(uint64_t x) const {
  uint64_t i = 0;
  __m256i cmp1 = _mm256_cmpgt_epi32(_mm256_loadu_si256((__m256i const*)summary),
                                    _mm256_set1_epi32(x));
  i = index_fs<32>(cmp1) - 1;
  x -= summary[i];
  i *= 16;
  __m256i cmp2 = _mm256_cmpgt_epi16(_mm256_loadu_si256((__m256i const*)(keys + i)),
                                    _mm256_set1_epi16(x));
  i += index_fs<16>(cmp2);
  return {i, i ? sum(i - 1) : 0};
}
\end{lstlisting}
The algorithm essentially uses parallel comparisons of $x$ against,
first, the summary and, then, in a specific segment.
The result of $\code{cmpgt\_epi16}(V, X)$ between the vectors $V$ and $X$
of 16-bit integers is
a vector $Y$ such that $Y[i]=2^{16}-1$ if $V[i]>X[i]$ and 0 otherwise.
From vector $Y$, we need to find the index of the ``first set'' (\code{fs})
16-bit field. This function, referred to as \code{index\_fs} in the code,
can be implemented efficiently using the built-in instruction
\code{ctz} (count trailing zeros) over 64-bit words.
(The function takes a template parameter representing the number of
bits of each integer in a 256-bit register.)
Some notes are in order.

Note that, since $\code{summary}[0]=0$, it is guaranteed that
\code{index\_fs(cmp1)} is always at least 1 (but not necessarily
true for \code{index\_fs(cmp2)}).

As already mentioned,
there is no \code{cmpgt} instruction for \emph{unsigned} integers in instruction
sets up to and including AVX2.
Nonetheless, observe again that the {\Search} algorithm shown here
is correct because we \emph{never} use the 15-th bit (sign) of the 16-bit integers
we load in the registers.

Lastly, observe that the {\Search} function returns a pair of integers,
not only the index $i$ but also the sum of the elements to the
``left of'' index $i$ for convenience, i.e., $\Sum(i-1)$ (or just 0 if $i=0$),
because this quantity is needed by the \code{select} operation in Figure~\ref{code:mbitmap}.


\parag{Other node types}
With \code{node128} it is, therefore, possible to handle bitmaps
of size up to $2^7 B$ bits.
We can now define a \code{node64} structure,
by applying the same design rules (and using similar algorithms)
as we have done for the \code{node128} structure.
The \code{node64} structure uses 8 segments, each holding 8 32-bit
keys. Its summary array holds 32-bit numbers too.
It follows that using a {\segtree} of height two, where
the root of the tree is a node of type \code{node64} and its children
are of type \code{node128}, it is possible ho handle bitmaps of size
up to $2^{6+7}B$ bits.
For a tree of height 3 we re-use again \code{node64}, for
up to $2^{6+6+7}B$ bits.
Adding another node, \code{node32},
made up of 4 64-bit values for the summary and
segment of 8 32-bit keys,
suffices to handle bitmaps of size up to $2^{24}B$ bits,
e.g., $2^{32}$ bits for $B=256$.
Table~\ref{tab:nodes}a summarizes these node types,
with Table~\ref{tab:trees}a showing how these
are used in the {\segtree}.

\begin{table}[t]
\centering
\caption{{\segtree} node types. The type \code{node}\emph{x} has fanout
equal to \emph{x}.}

\subfloat[AVX2]{
\scalebox{0.9}{\input{tables/nodes_avx2.tex}}
}
\subfloat[AVX-512]{
\scalebox{0.9}{\input{tables/nodes_avx512.tex}}
}

\label{tab:nodes}
\end{table}

\begin{table}[t]
\centering
\caption{{\segtree} configurations used to index a bitmap with
blocks of $B$ bits, $B < 2^{11}$.}

\subfloat[AVX2]{
\scalebox{0.9}{\input{tables/trees_avx2.tex}}
}
\subfloat[AVX-512]{
\scalebox{0.9}{\input{tables/trees_avx512.tex}}
}

\label{tab:trees}
\end{table}

\subsection{Using AVX-512}
Using registers of 512 bits clearly increases the
parallelism of both updates and searches as
$2\times$ more integers can be packed in a register
compared to AVX2 that uses 256-bit registers.
We can again define three node types to make full exploitation
of the new AVX-512 instruction set:
\code{node512}, \code{node256}, and \code{node128}.
Their design is akin to that for AVX2 in the previous section,
with details given in Table~\ref{tab:nodes}b.
Table~\ref{tab:trees}b shows, instead, that the
height of the {\segtree} can be reduced by 1 compared to
AVX2.
We also expect this to help in lowering the runtime
as cache misses are reduced.

Another (minor) advantage of AVX-512 over AVX2 is that
instructions implementing comparisons directly return
a mask vector of 8, 16, 32, or 64 bits,
instead of a wider SIMD register.
This resulting mask can then be used as input for
another instrinsic, such as \code{ctz}.
In our case, the function \code{index\_fs} used
during a search operation
just translates to one single \code{ctz} call
(instead of 4 calls as in the case for a 256-bit
register output by \code{\_mm256\_cmpgt}).

The only \emph{current} limitation of AVX-512 is that is not
as widespread as AVX2, so algorithms based on such
instruction set are less portable.

\subsection{Experimental Analysis}
In this section, we measure the runtime of {\Sum}, {\Update}, and {\Search},
for the introduced {\segtree} data structure and
optimized state-of-the-art approaches, i.e.,
the {\fentree}-based data structures described by~\citet*{marchini2020compact},
and implemented in the \textsf{Sux}~\cite{sux} C++ library.
We reuse the nomenclature adopted by the authors
in their own paper:
\textsf{Byte} indicates compression at the byte granularity;
\textsf{Fixed} indicates no compression;
\textsf{F} stands for classic Fenwick layout;
and \textsf{L} stands for level-order layout.
For these experiments, we used arrays generated at random of size $2^k$ integers,
for $k=8..24$.
Since the runtime of the operations on the {\segtree}
does \emph{not} depend on the block size $B$,
we fix $B$ to 256 in these experiments.
Therefore, each integer of the input array
is a random value in $[0..256]$.
(Each element of the array logically models the population
count of a 256-bit block, so that the maximum array size cannot exceed
$2^{24}$ for a bitmap of size $2^{32}$ bits.)

\begin{figure}[!t]
\centering
\includegraphics[width=140mm]{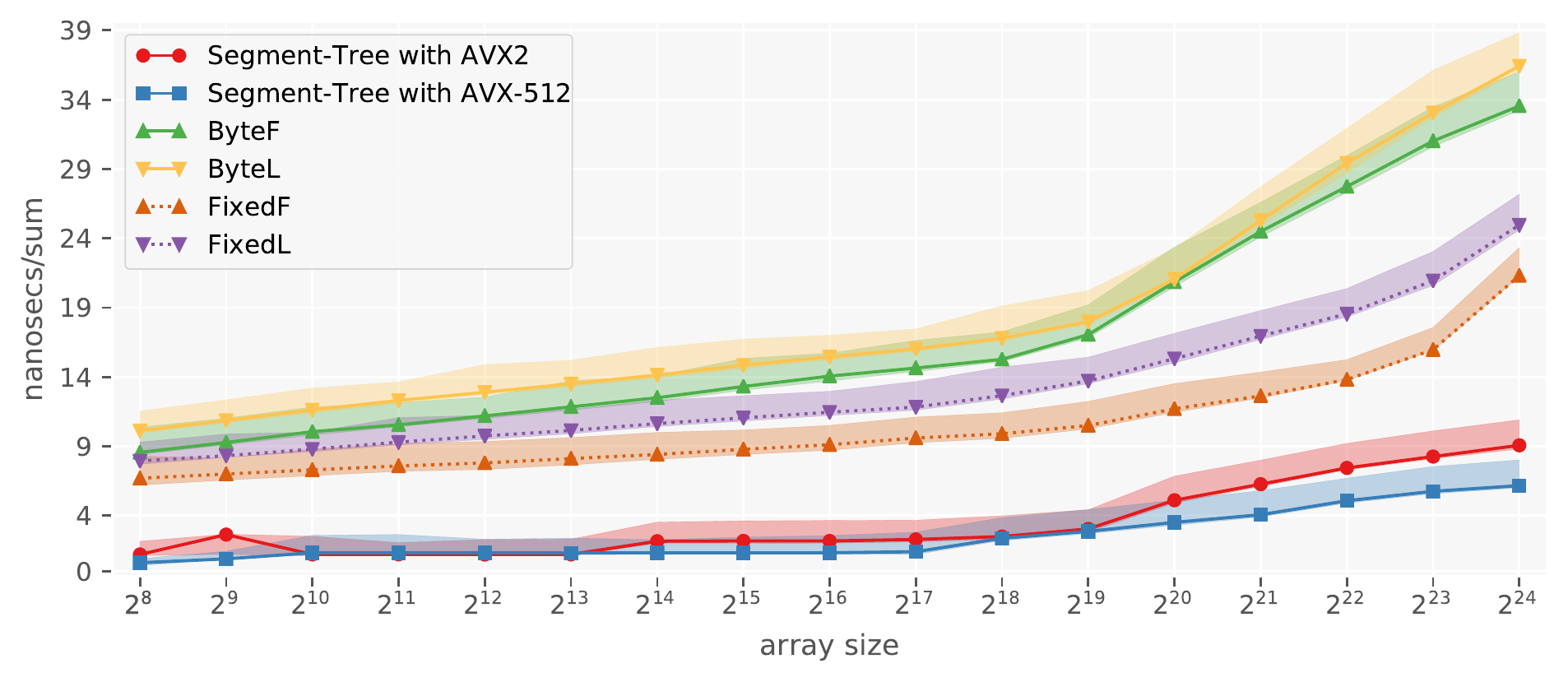}
\includegraphics[width=140mm]{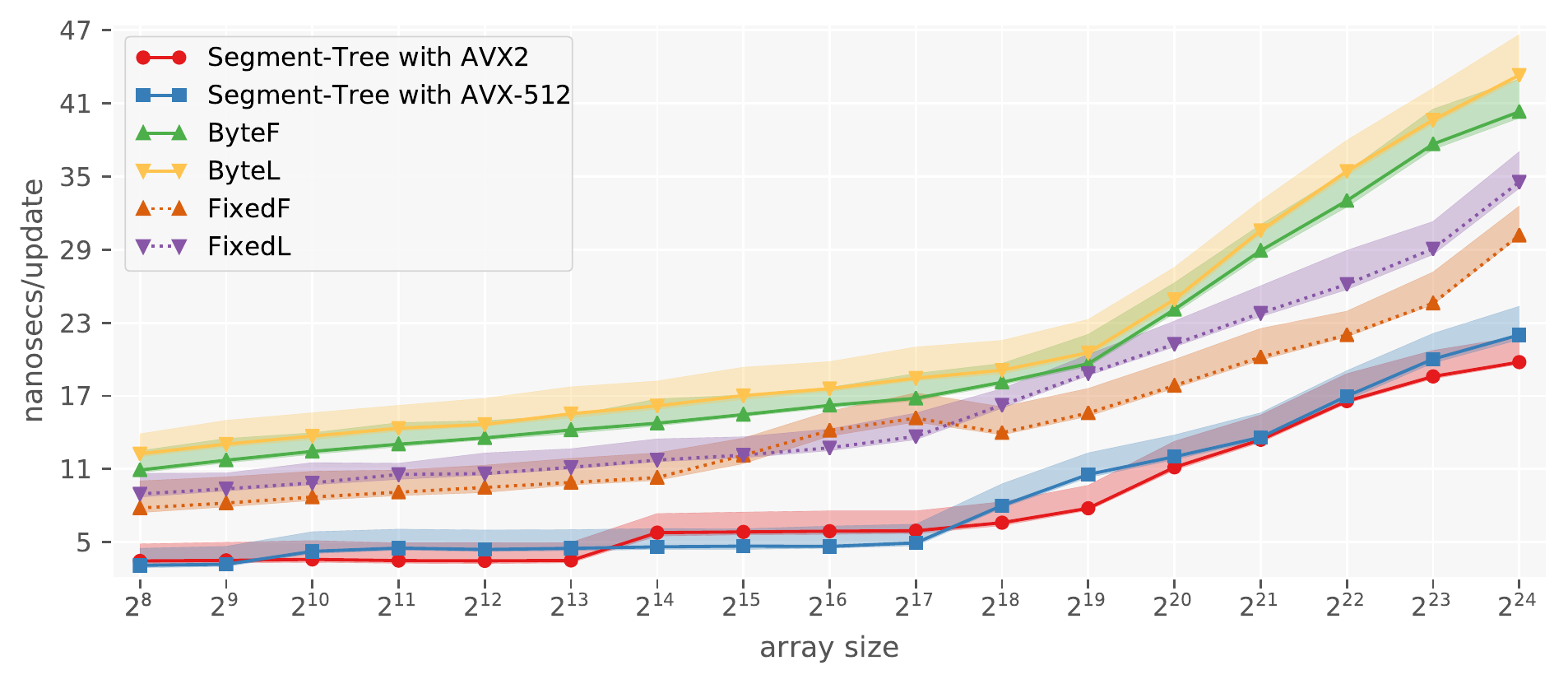}
\includegraphics[width=140mm]{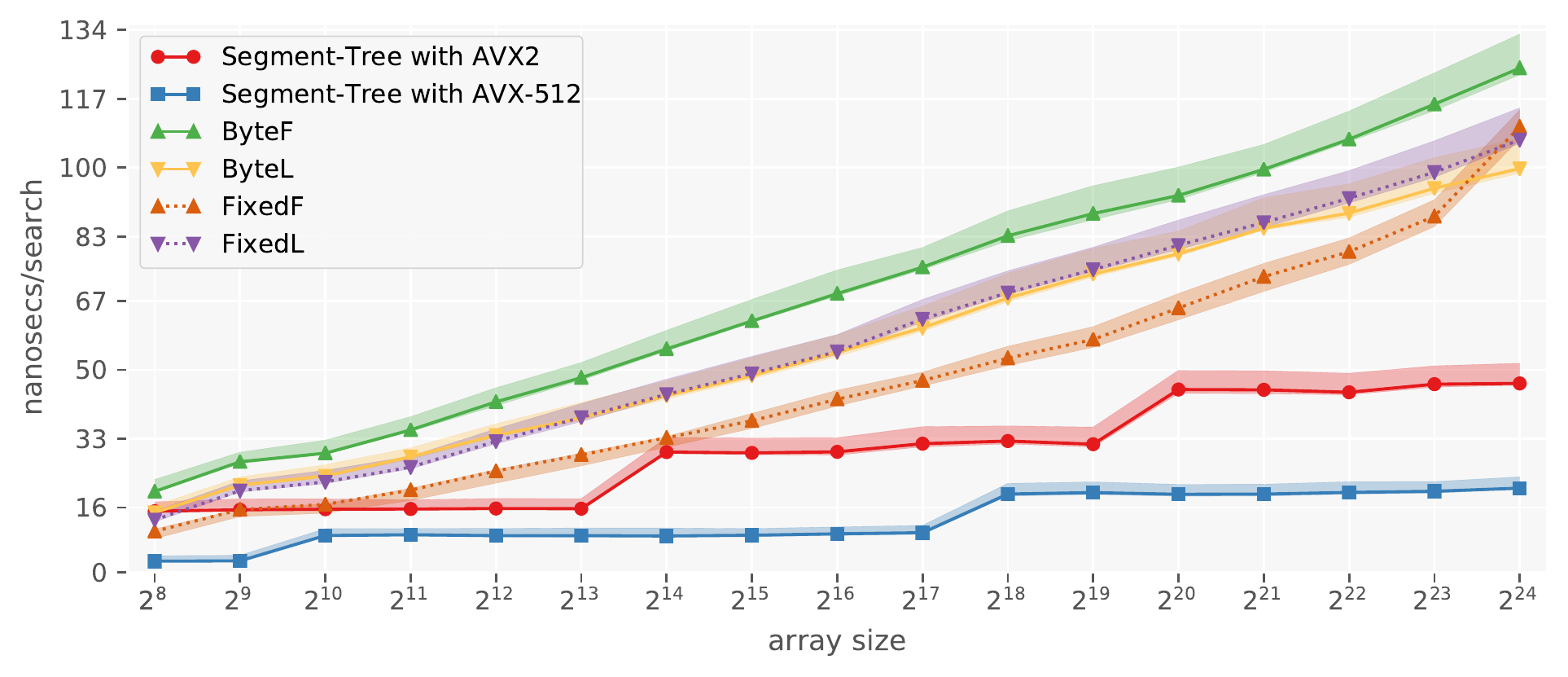}
\caption{Average nanoseconds spent per operation by
searchable prefix-sum data structures.}
\label{fig:sum_update_search}
\end{figure}


Figure~\ref{fig:sum_update_search} illustrates the result of the
experiments.
We shall first consider the {\segtree} data structure described in this
section and compare the use of AVX2 with AVX-512.
The runtime of {\Sum} and {\Update} is very similar
for both versions of the {\segtree}.
Recall that every tree operation translates into
some of the corresponding operations performed at the nodes --
one such operation per level of the tree.
For {\Sum} and {\Update}, these node operations are
\emph{data-independent} and our implementation exploits this
fact to leave the processor pipeline the opportunity to execute
them out of order. Therefore, only one level of difference
between the two tree heights
does not suffice to make the two curves sufficiently distinct.

This consideration does \emph{not} apply, instead, to the {\Search}
operation because the result of {\Search} at level $\ell$
is used to perform {\Search} at level $\ell+1$, hence
making the {\Search} algorithm inherently sequential.
As a consequence of this fact,
observe that the runtime of {\Search} is substantially
higher than that of {\Sum} and {\Update} (not only because of
the more complicated algorithm).
In this case, the use of AVX-512 shines over AVX2
-- often by a factor of $2\times$ --
because one less level to traverse makes a big difference
in the runtime of a sequential algorithm.
(Also notice how both curves rise when one more level of the tree is in use.)

Compared to solutions based on the {\fentree},
the {\segtree} is consistently faster and by a good margin, confirming
that the joint use of branch-free code
and SIMD instructions have a significant impact on the
practical performance of the data structures.

The space of the {\segtree} is slightly more than the space needed to represent
the input array itself, assuming that a fixed number of bytes is used to
store each element.
For our choice of block size,
each array element can be stored in 2 bytes, so the
number of bytes spent by the {\segtree} for each element
will be slightly more than 2 for sufficiently large inputs, i.e.,
$\approx$2.29 for AVX2 and $\approx$2.13 for AVX-512.
In fact, the overhead of the tree hierarchy progressively vanishes
as the tree height increases because nodes with large fanout are used.
(It is easy to derive the \emph{exact} number of bytes consumed
by the data structure from the tree height and the space of each
node type reported in Table~\ref{tab:nodes}.)
This ultimately means that, when the {\segtree} is used as a bitmap index,
its extra space is just
$(2.29 \times 8)/256 \times 100\% < 7.2\%$ of that of the bitmap
itself with blocks of 256 bits,
and less than 3.6\% with blocks of 512 bits.

The compressed {\fentree} data structures perform similarly
to the {\segtree} (the space for \textsf{ByteF} and \textsf{ByteL}
is the same), taking $\approx$2 bytes per element.
The \textsf{Fixed} variants take considerably more space
because every element is represented with 4 or 8 bytes.

In conclusion,
the searchable prefix-sum data structure described in this section
-- a {\segtree} with SIMD-ized updates and searches --
consumes space comparable to that of a compressed {\fentree}
while being significantly faster.



%% file: tables/nodes_avx2.tex
\begin{tabular}{c c c c}
\toprule
Type 
& Summary & Segment & Bytes \\
\midrule

\code{node32}  & 4 $\times$ 64-bit & \,\,8 $\times$ 32-bit & 160 \\
\code{node64}  & 8 $\times$ 32-bit & \,\,8 $\times$ 32-bit & 288 \\
\code{node128} & 8 $\times$ 32-bit &    16 $\times$ 16-bit & 288 \\

\bottomrule
\end{tabular}

%% file: tables/nodes_avx512.tex
\begin{tabular}{c c c c}
\toprule
Type & Summary & Segment & Bytes \\
\midrule

\code{node128}  & \,\,8 $\times$ 64-bit & 16 $\times$ 32-bit &  576 \\
\code{node256}  &    16 $\times$ 32-bit & 16 $\times$ 32-bit & 1088 \\
\code{node512}  &    16 $\times$ 32-bit & 32 $\times$ 16-bit & 1088 \\

\bottomrule
\end{tabular}

%% file: tables/trees_avx2.tex
\begin{tabular}{c c c}
\toprule
Tree height & Node hierarchy & Bitmap size \\
\midrule

4 & \code{node32}  & up to $2^{24}B$ bits \\
3 & \code{node64}  & up to $2^{19}B$ bits \\  
2 & \code{node64}  & up to $2^{13}B$ bits \\
1 & \code{node128} & up to $2^{7} B$ bits \\

\bottomrule
\end{tabular}

%% file: tables/trees_avx512.tex
\begin{tabular}{c c c}
\toprule
Tree height & Node hierarchy & Bitmap size \\
\midrule

3 & \code{node128} & up to $2^{24}B$ bits \\  
2 & \code{node256} & up to $2^{17}B$ bits \\
1 & \code{node512} & up to $2^{9}B$ bits \\

\bottomrule
\end{tabular}

%% file: small_bitmaps.tex
\section{Rank/Select Queries over Small Bitmaps}\label{sec:small_bitmaps}

In this section we study the problem of answering {\Rank} and {\Select} queries
over small bitmaps of $B = 256$ and 512 bits
(four and eight 64-bit words respectively)
\emph{without} auxiliary space.
We first describe the algorithms under a unifying implementation framework;
then discuss experimental results to choose the fastest algorithms.

\begin{table}[t]
\renewcommand\thesubtable{\arabic{subtable}}
\centering
\caption{The two-step framework for {\Rank} with approaches to perform
the steps of (1) popcount and (2) prefix-sum.
}
\subfloat[popcount]{
\scalebox{1.0}{\input{tables/counting_rank.tex}}
\label{tab:counting_rank}
}
\hspace{4mm}
\subfloat[prefix-sum]{
\scalebox{1.0}{\input{tables/summing_rank.tex}}
\label{tab:summing_rank}
}
\label{tab:rank}
\end{table}

\begin{figure}[tb]
\begin{lstlisting}
__m256i prefix_sum_256(__m256i C) {
  static const int idx1 = _MM_SHUFFLE(2, 1, 0, 3);
  static const int idx2 = _MM_SHUFFLE(1, 0, 3, 2);
  C = _mm256_add_epi64(C, _mm256_maskz_permutex_epi64(0b11111110, C, idx1));
  C = _mm256_add_epi64(C, _mm256_maskz_permutex_epi64(0b11111100, C, idx2));
  return C;
}

__m512i prefix_sum_512(__m512i C) {
  static const __m512i idx1 = _mm512_set_epi64(6, 5, 4, 3, 2, 1, 0, 7);
  static const __m512i idx2 = _mm512_set_epi64(5, 4, 3, 2, 1, 0, 7, 6);
  static const __m512i idx3 = _mm512_set_epi64(3, 2, 1, 0, 7, 6, 5, 4);
  C = _mm512_add_epi64(C, _mm512_maskz_permutexvar_epi64(0b11111110, idx1, C));
  C = _mm512_add_epi64(C, _mm512_maskz_permutexvar_epi64(0b11111100, idx2, C));
  C = _mm512_add_epi64(C, _mm512_maskz_permutexvar_epi64(0b11110000, idx3, C));
  return C;
}
\end{lstlisting}
\caption{A SIMD-based implementation of the parallel prefix-sum algorithm described by~\citet*{hillis1986data},
for 256 and 512 bits.
It leverages the instructions
\code{maskz\_permutex} and \code{maskz\_permutexvar} available
in the instruction set AVX-512.
\label{code:prefixsum_avx512}}
\end{figure}

\subsection{Rank}
A common framework to solve {\Rank} for an index $0 \leq i < B$
consists of the following two steps.
\begin{enumerate}
\item \emph{Count} the numbers of ones 
appearing in the first
$j = \floor{i / 64}$ words, where the $j$-th word is appropriately masked
to zero the most significant $64 - (i+1) \bmod 64$ bits.
\item \emph{Prefix-sum} the resulting counters.
\end{enumerate}
From a logical point of view, the result of step (1) is an array $C[0..B/64)$
of counters so that the final result for $\Rank(i)$ is computed
as $C[0] + \dots + C[j]$ in step (2).

Several algorithms to perform these two steps are summarized in Table~\ref{tab:rank}.
By combining such basic steps we obtain different implementations
of {\Rank}.

We have three different approaches to implement step (1).
The first is the broadword algorithm by~\citet*[Algo. 1]{vigna2008broadword}.
The second is the \code{\_\_builtin\_popcountll} {instruction}
available from SSE4.2.
The third is a parallel algorithm using SIMD developed by~\citet[Figure 10]{mula2017faster},
to concurrently compute 4 or 8 popcount results, i.e.,
filling the vector $C$.
(We took the original implementation made available by the authors at \url{https://github.com/WojciechMula/sse-popcount}.)

Also for step (2) we have three different approaches.
The first is a simple loop that incrementally sums $C[k]$ for $0 \leq k \leq j$,
and requires a test for $j$ at each iteration.
The second is an unrolled loop that computes all the prefix sums
$C[k] = C[k] + C[k-1]$ for $0 < k < B/64$ and returns $C[j]$.
The approach does not require any branch.
The third is a SIMD-based implementation of the parallel prefix-sum algorithm devised by~\citet*{hillis1986data}.
In Figure~\ref{code:prefixsum_avx512} we show the two implementations
for, respectively, 256 and 512 bits.
Note that the last two approaches work in a branch-free manner
but need to popcount \emph{all} the $B/64$ words.

\begin{figure}[!t]
\centering
\includegraphics[width=140mm]{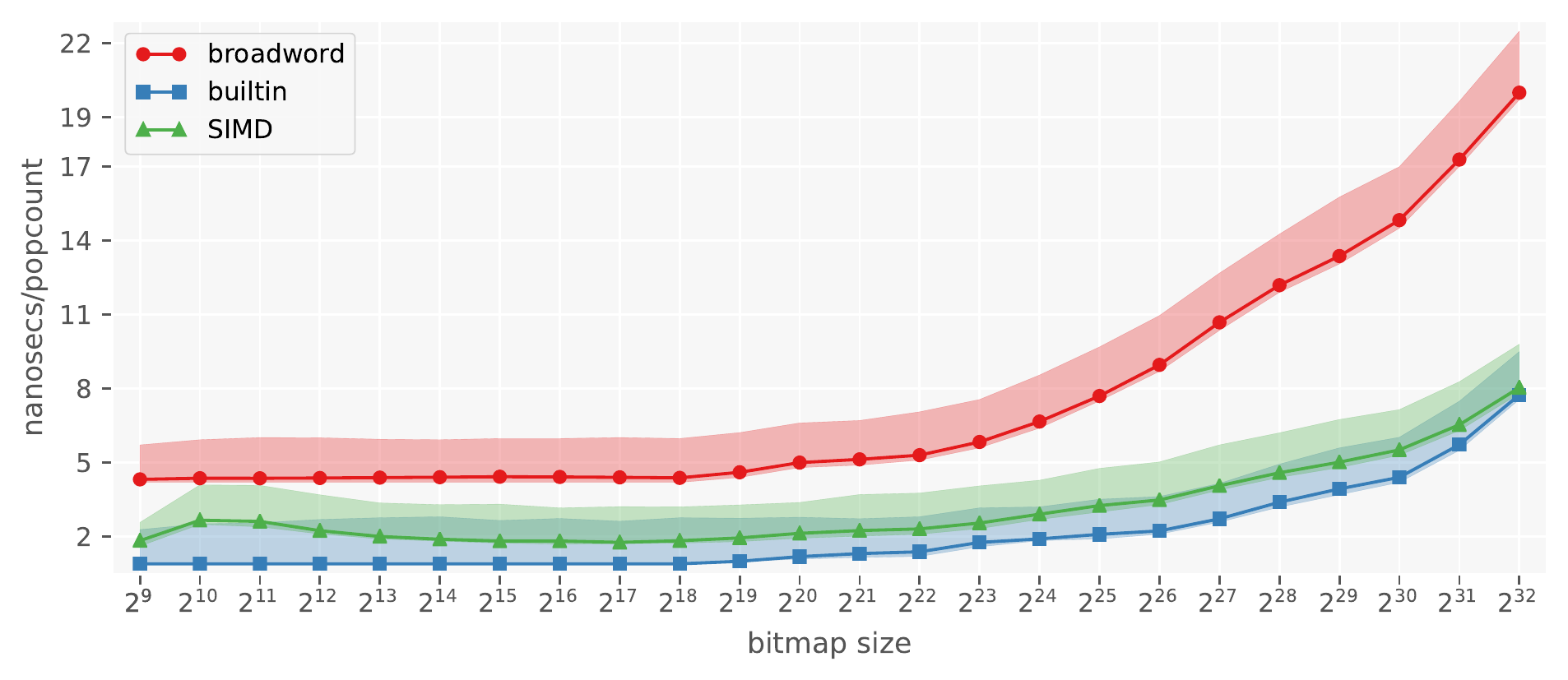}


\caption{Average nanoseconds spent per pop-count (step (1) in Table~\ref{tab:rank})
by different algorithms, for $B=256$.
The trend for $B=512$ is the same, but the gap between broadword and
the other approaches is more evident.
}

\label{fig:popcount}
\end{figure}

\begin{figure}[!t]
\centering
\subfloat[$B = 256$ bits]{
\includegraphics[width=140mm]{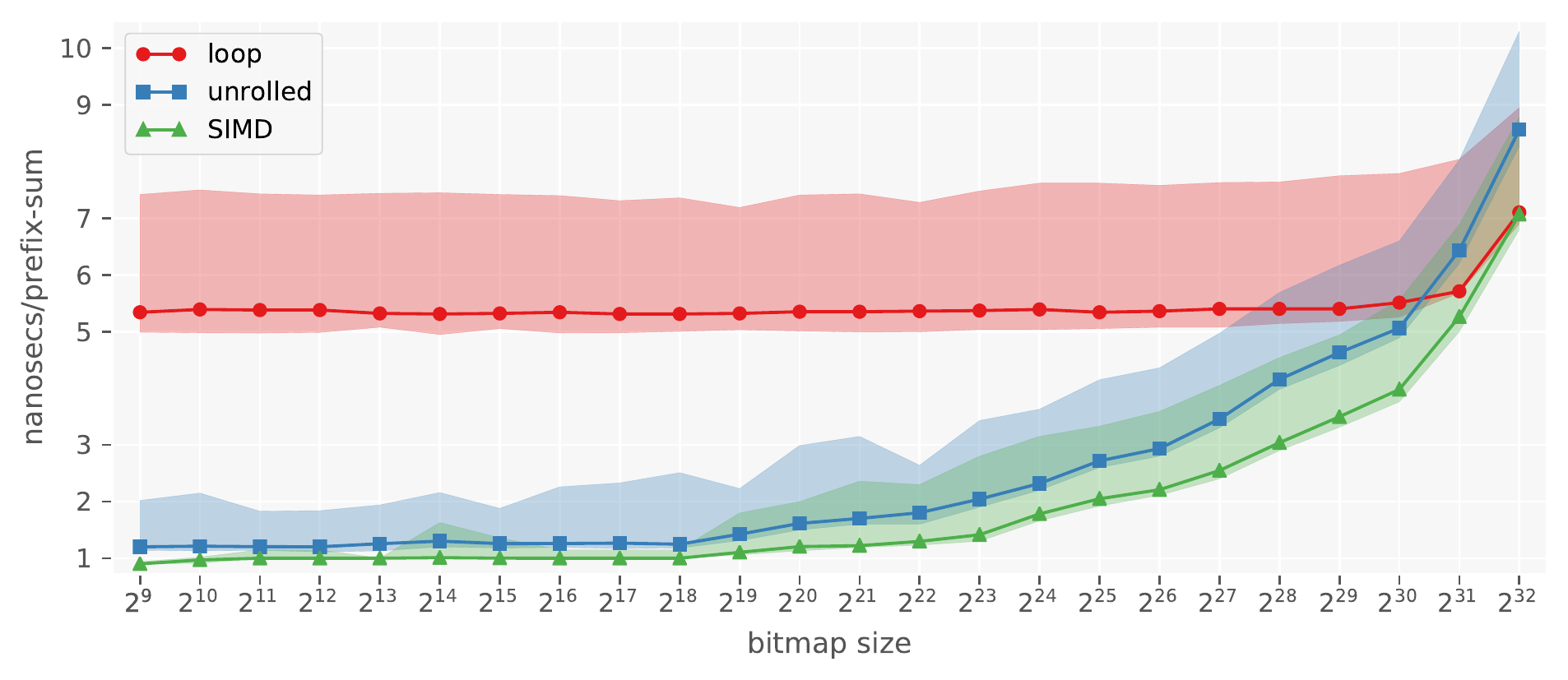}
}

\subfloat[$B = 512$ bits]{
\includegraphics[width=140mm]{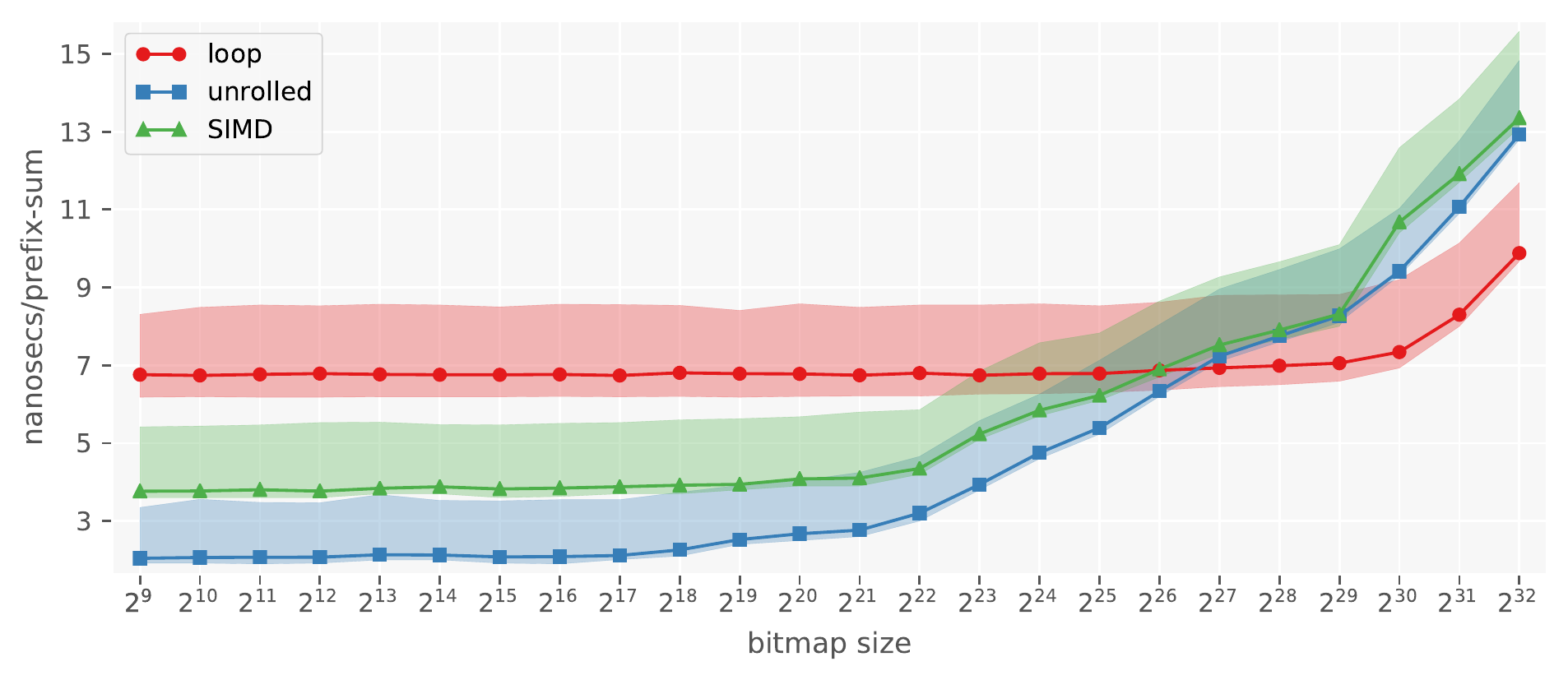}
}

\caption{Average nanoseconds spent per prefix-sum (step (2) in Table~\ref{tab:rank})
by different algorithms.}
\label{fig:prefixsum}
\end{figure}

\begin{figure}[!t]
\centering
\subfloat[$B = 256$ bits]{
\includegraphics[width=140mm]{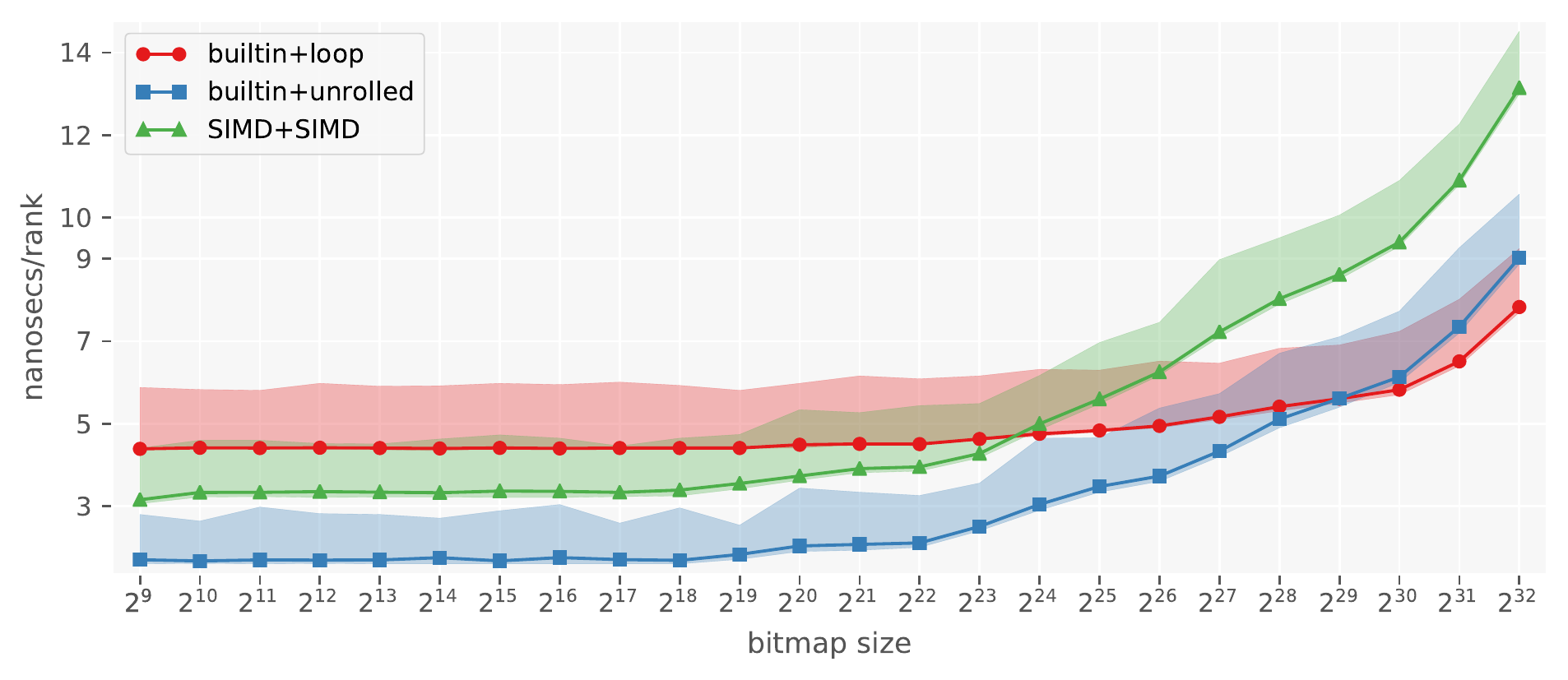}
}

\subfloat[$B = 512$ bits]{
\includegraphics[width=140mm]{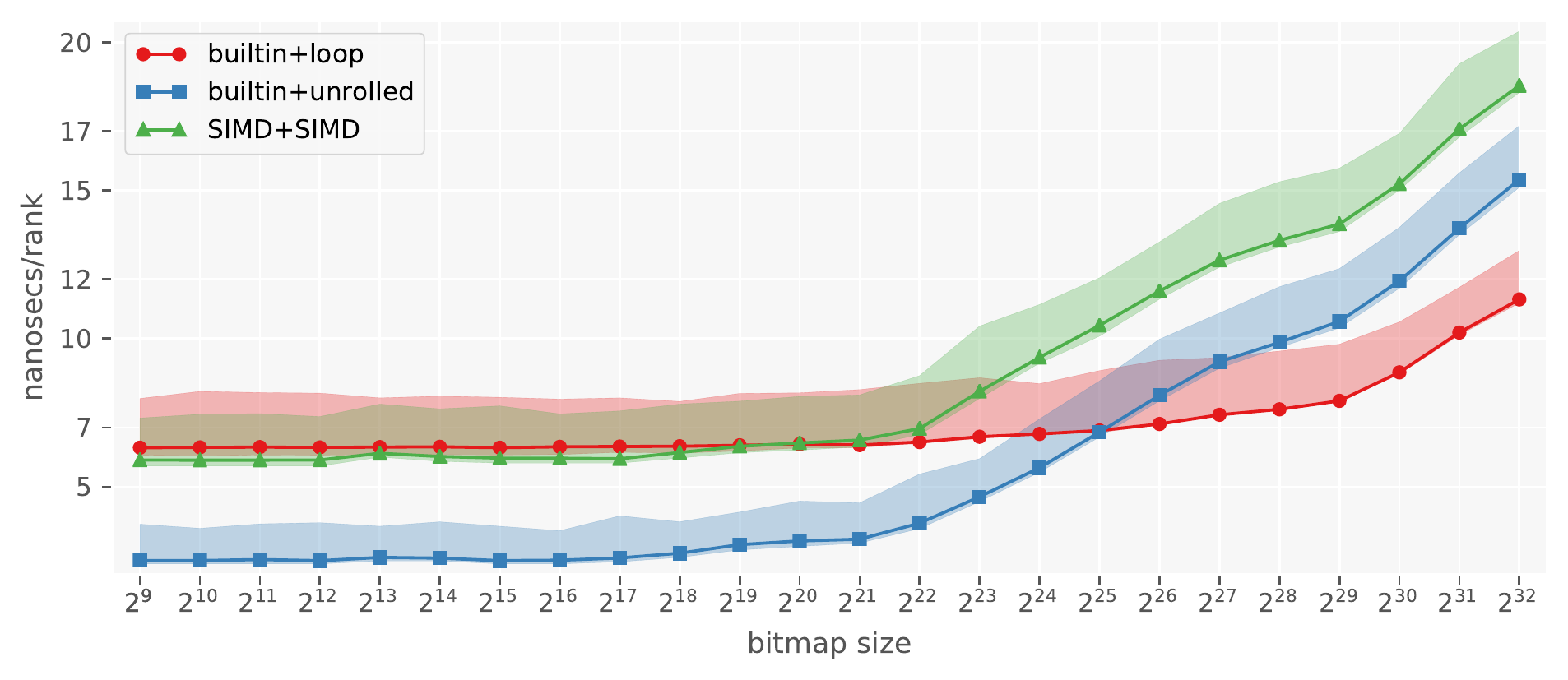}
}
\caption{Average nanoseconds spent per {\Rank} by different algorithms.}
\label{fig:rank}
\end{figure}

We now consider the runtimes achieved by the various approaches
reported in Table~\ref{tab:rank}.
For all such experiments, we generated random bitmaps of increasing size
and random queries of the form $\langle j,i_j \rangle$,
indicating that a given algorithm has to run
over block $j$ and with index $i_j$ as input parameter.

Figure~\ref{fig:popcount} shows the runtimes for step (1).
The fastest approach is to use the built-in popcount
instruction, with the SIMD-based algorithm catching up
(especially on the larger $B=512$). The broadword algorithm
is outperformed by the other two approaches.
Concerning step (2), Figure~\ref{fig:prefixsum} shows that the
use of an unrolled loop provides consistently better results
than a traditional loop
for a wide range of practical bitmap sizes,
because it completely avoids branches and increases the
instruction throughput.
It is comparable with (for $B=256$) or even better than SIMD
(for $B=512$).
However, when the size of a bitmap is very large, using a simple loop
is faster.
This is because large bitmaps involve slower memory accesses,
resulting in the memory latency dominating the cost of the CPU pipeline flush.
In that case, the branchy implementation becomes faster since unnecessary
memory accesses are avoided.
(In fact, the difference is more evident for the case $B=512$
rather than for $B=256$.)
This is consistent with prior work~\cite{pibiri2020practical}.

Lastly, in the light of the results in Figure~\ref{fig:popcount} and~\ref{fig:prefixsum},
we combine some of the approaches to obtain
the fastest {\Rank} algorithm.
Our expectation is that the combination \textsf{builtin+unrolled}
performs best until the bitmap size becomes very large,
when eventually the combination \textsf{builtin+loop} wins out.
Also the implementation \textsf{SIMD+SIMD} is expected to be competitive
with, but actually worse than, these two identified combinations.
Figure~\ref{fig:rank} shows the runtimes of such algorithms and
confirms our expectations.

\subsection{Select}
A common framework to solve {\Select} for an index $0 \leq i < B$ consists
of the following three steps.
\begin{enumerate}
\item \emph{Count} the numbers of ones appearing in each word, using an array $C[0..B/64)$.
\item \emph{Prefix-sum} $C$ and \emph{search} for the $j$-th word, $w_j$,
containing the $i$-th bit set.
\item \emph{Compute} {\Select} locally in $w_j$, an operation that
we denote by {\SelectW} (select-in-word).
\end{enumerate}
More formally,
step (2) determines the smallest $j$ such that $i < C[j]$ and
step (3) returns $j64 + \SelectW(w_j, i - C[j-1])$.
Approaches to perform these steps are summarized in Table~\ref{tab:select}.

\begin{table}[!t]
\renewcommand\thesubtable{\arabic{subtable}}
\centering
\caption{The three-step framework for {\Select} with approaches to perform
the steps of (1) popcount, (2) search, and (3) select-in-word.
}

\subfloat[popcount]{
\begin{tabular}{c}
\toprule
same as in Table~\ref{tab:rank} \\
\bottomrule
\end{tabular}
}
\hspace{4mm}
\subfloat[search]{
\scalebox{1.0}{\input{tables/searching_select.tex}}
\label{tab:searching_select}
}
\hspace{4mm}
\subfloat[select-in-word]{
\scalebox{1.0}{\input{tables/selecting_in_word.tex}}
\label{tab:selecting_in_word}
}
\label{tab:select}
\end{table}

For step (1), we reuse the three approaches in Table~\ref{tab:rank}
adopted for {\Rank}.
For step (2), we have two approaches:
the first is a simple loop that incrementally computes
the prefix-sums in $C$ and checks if $i < C[j]$;
the second is a branch-free implementation that computes the prefix-sums in parallel and finds the target $j$-th word using SIMD AVX-512 instructions.
For step (3), we have three approaches.
Two approaches are based on broadword programming,
developed by~\citet*{vigna2008broadword} and~\citet*{gog2014optimized},
respectively.
We used the implementations available from the
libraries \textsf{Succinct}~\cite{grossi2013design}
and \textsf{Sdsl}~\cite{gog2014optimized,gog2014theory}.
(We took advantage of built-in instructions whenever possible.
This makes both algorithms faster compared to when intrinsics are not used.)
The third approach is a  one-line algorithm using the
\emph{parallel bits deposit} (pdep) intrinsic devised by~\citet{pandey2017fast}.

\begin{figure}[!t]
\centering
\includegraphics[width=140mm]{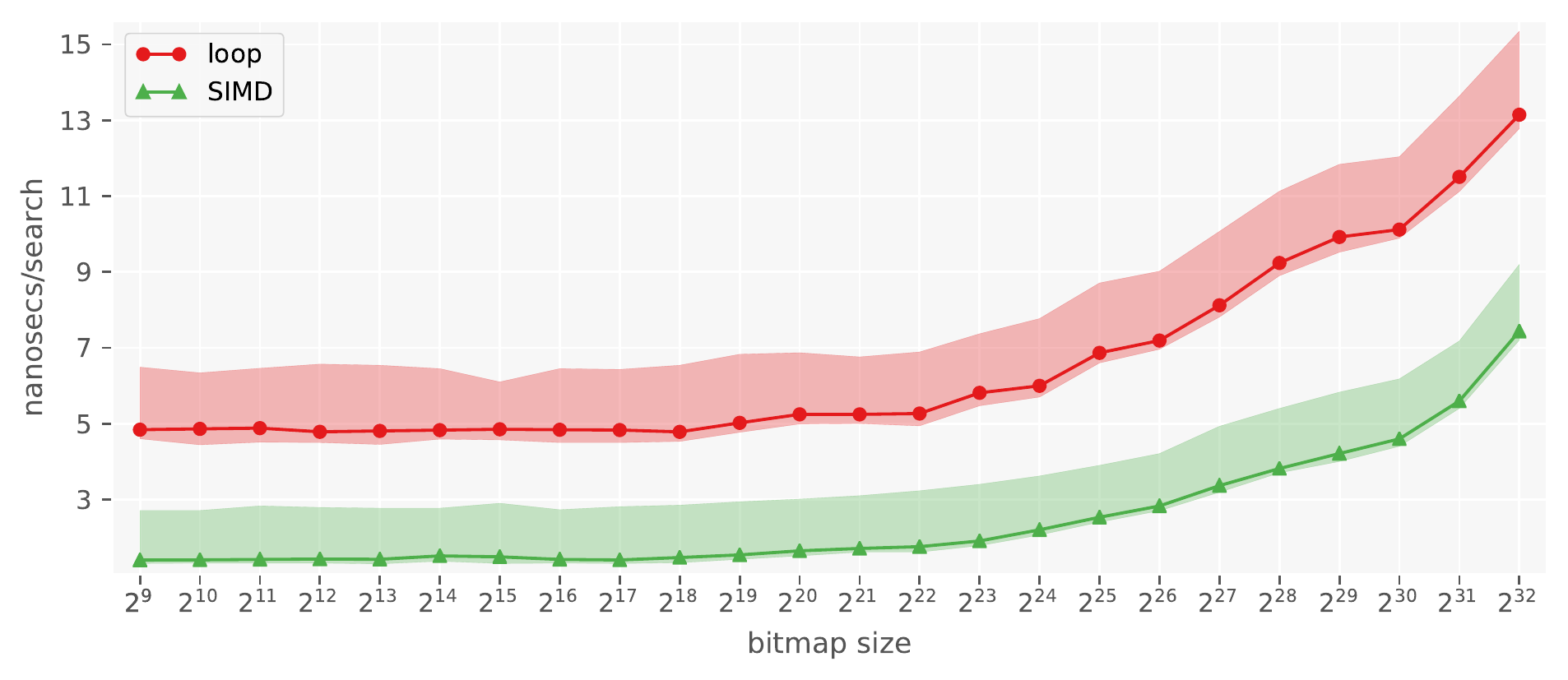}

\caption{Average nanoseconds spent per search (step (2) in Table~\ref{tab:select})
by different algorithms, for $B=256$.
(The shape for $B=512$ is the same.)
}
\label{fig:search}
\end{figure}

\begin{figure}[!t]
\centering
\includegraphics[width=140mm]{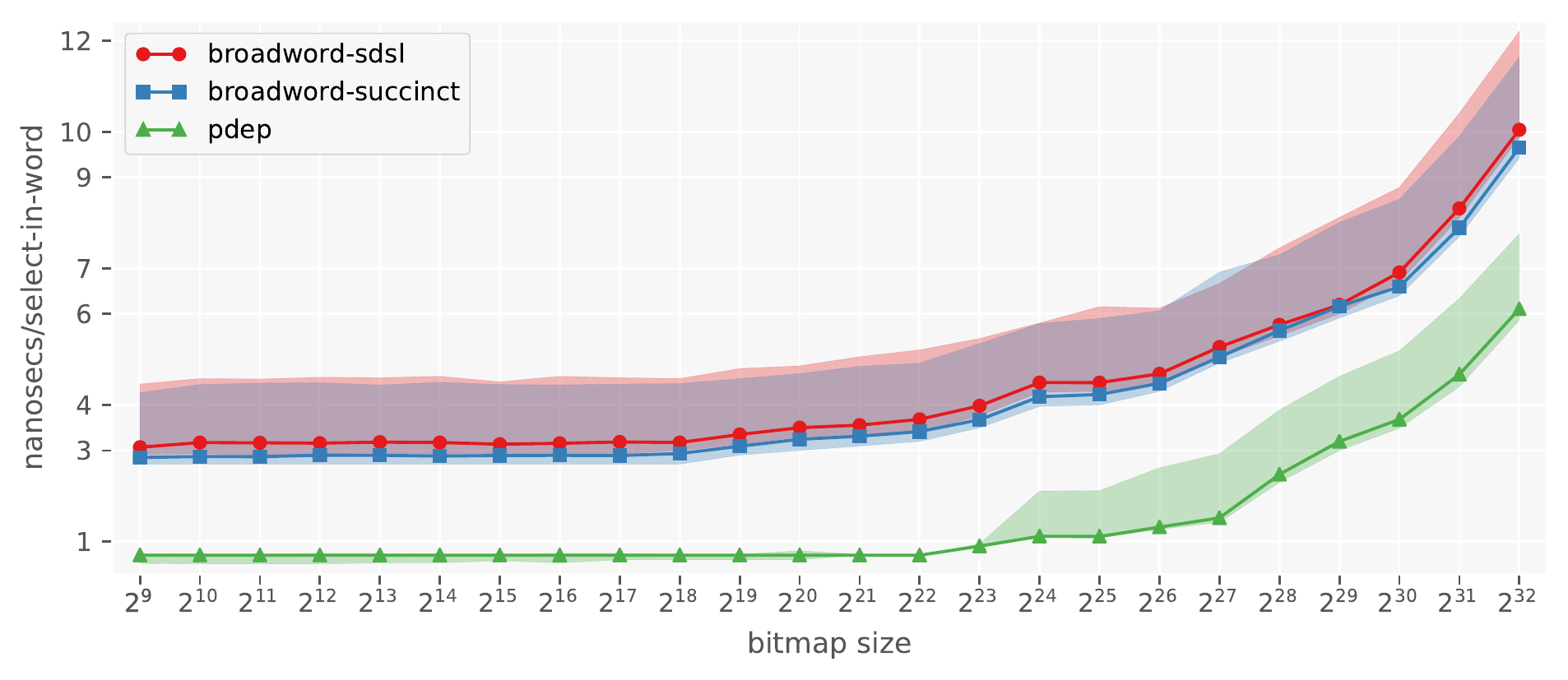}
\caption{Average nanoseconds spent per select-in-word (step (3) in Table~\ref{tab:select})
by different algorithms.}
\label{fig:select64}
\end{figure}

To benchmark the approaches in Table~\ref{tab:select},
we reuse the same methodology adopted for {\Rank}:
random bitmaps of increasing size under a random workload
of $\langle j,i_j \rangle$ queries.
Answering $\Select(i_j)$ over block $j$ makes the runtime
practically independent from
the density of the \bit{1}s, which we fixed to 30\%
in these experiments.

From the results in Figure~\ref{fig:search} and~\ref{fig:select64},
we see that SIMD is very effective in implementing the search step
(which is also coherent with the results in Section~\ref{sec:prefix_sums})
and the use of pdep is much faster than broadword approaches
(as also noted in the paper by~\citet{pandey2017fast}).



\begin{figure}[!t]
\centering
\subfloat[$B = 256$ bits]{
\includegraphics[width=140mm]{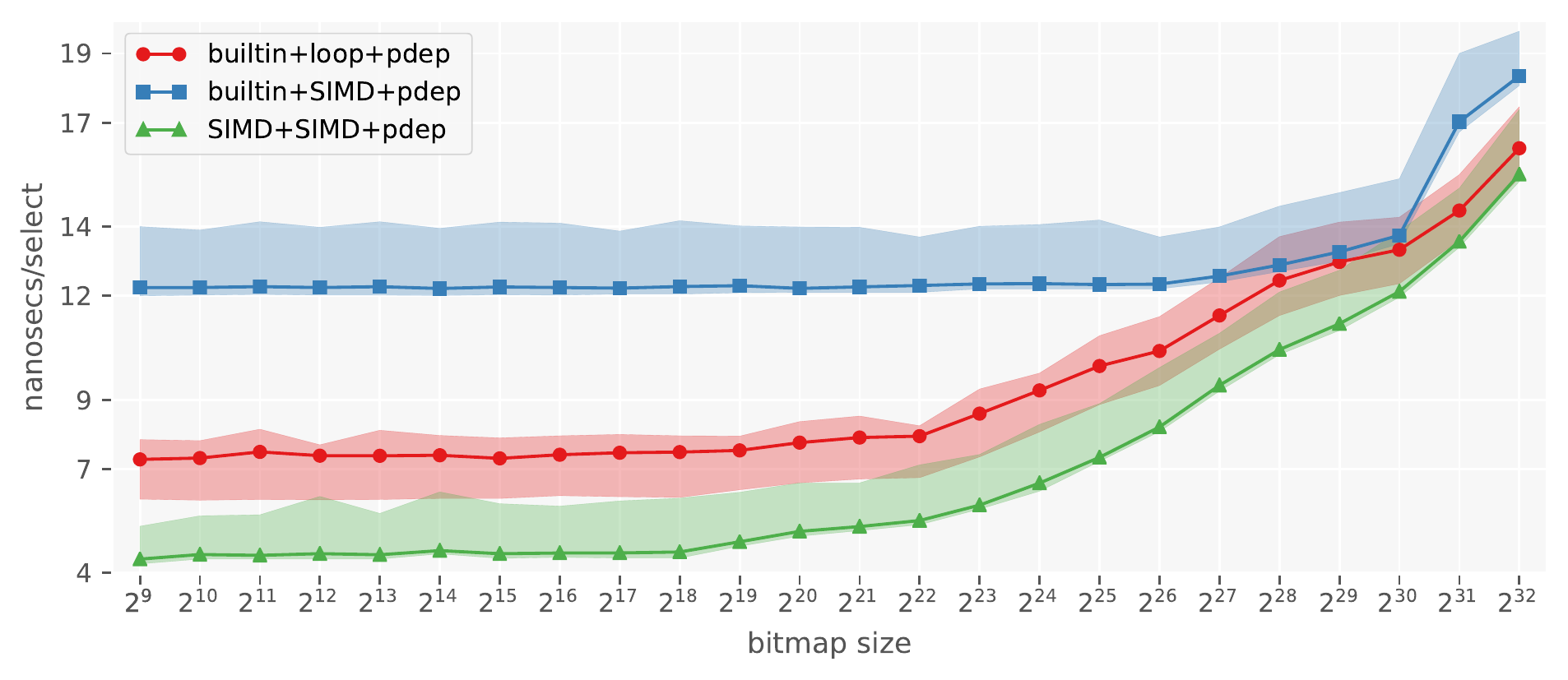}
}

\subfloat[$B = 512$ bits]{
\includegraphics[width=140mm]{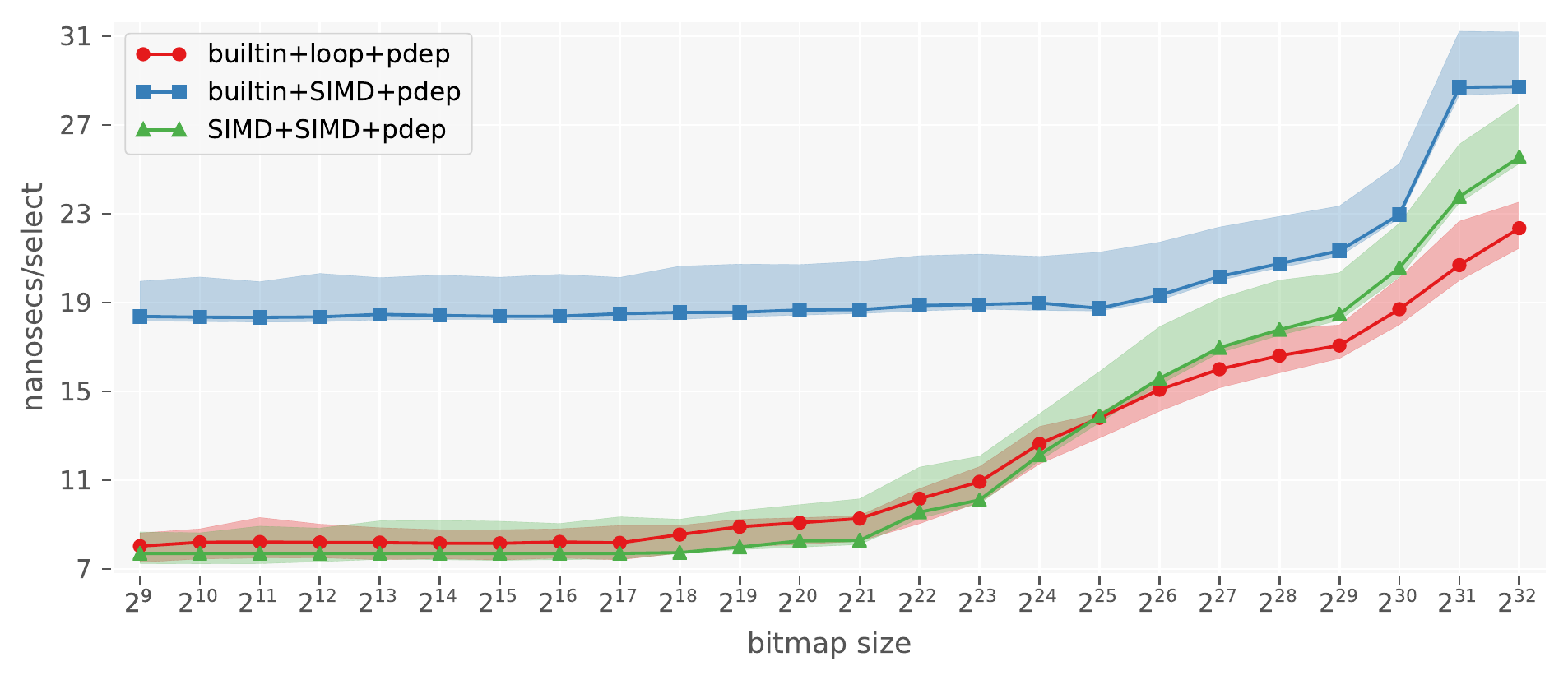}
}

\caption{Average nanoseconds spent per {\Select} by different algorithms.}
\label{fig:select}
\end{figure}

As similarly done for {\Rank}, we now discuss some combinations
of the basic steps
to identify the fastest {\Select} algorithm.
Since pdep is always the fastest at performing in-word-selection,
we exclude broadword approaches.
Clearly, we expect that a combination of \textsf{SIMD+pdep} for the
last two steps performs best.
The final result shown in Figure~\ref{fig:select} again meets our expectations
as the combination \textsf{SIMD+SIMD+pdep} is almost always the fastest.
However, note that also the combination \textsf{builtin+loop+pdep}
is good (and even the best for $B=512$ and large bitmaps)
because the first two steps are inherently merged into one, i.e.,
popcount is performed within the searching loop, whereas
the use of SIMD for step (2) requires computing popcount for all words.
As already noticed, this performs fewer memory accesses that
are very expensive for large bitmaps.


\subsection{Conclusions}

In the light of the experimental analysis in this section of the paper,
we now summarize the two fastest algorithms for {\Rank} and {\Select}.
\begin{itemize}
\item For {\Rank}, use \textsf{builtin+unrolled} for all bitmaps
of size up to $2^{25}$ bits (as a rule of thumb),
and switch to \textsf{builtin+loop} for larger bitmaps.
\item For {\Select}, use \textsf{SIMD+SIMD+pdep} if AVX-512 instructions are
available, or \textsf{builtin+loop+pdep} otherwise.
\end{itemize}
We recall that we are going to use these selected algorithms
over the blocks of a mutable bitmap data structure,
as exemplified in Figure~\ref{code:mbitmap}.

Now we report how much the runtime increases,
in percentage,
when considering these algorithms with $B=512$
compared to the case where $B=256$.
To do so, we compute the \emph{average} increase for three intervals of bitmap
size $u$, that model small, medium, and large bitmaps respectively:
(i)   $2^9 \leq u < 2^{21}$,
(ii)  $2^{21} \leq u < 2^{25}$, and
(iii) $2^{25} \leq u \leq 2^{32}$.
For {\Rank}, we determined an average increase of (i) 50\%, (ii) 80\%, and (iii) 40\%.
For {\Select}, we obtained (i) 80\%, (ii) 80\%, and (iii) 50\%.
These numbers show the doubling the block size $B$ sometimes causes
the runtime to \emph{nearly} double (+80\%) as one would expect,
but the increase is only 40-50\% more on large bitmaps.

Lastly, we also conclude that
SIMD instructions are more effective for {\Select} rather than {\Rank}.
This is because
potentially \emph{all} words in the block must be searched, differently
than {\Rank} that does not need any search,
and this can be done quickly with SIMD.

%% file: tables/counting_rank.tex
\begin{tabular}{l}
\toprule
broadword~\cite[Alg. 1]{vigna2008broadword} \\
built-in instruction \\
SIMD~\cite[Figure 10]{mula2017faster} \\
\bottomrule
\end{tabular}

%% file: tables/summing_rank.tex
\begin{tabular}{l}
\toprule
loop \\
unrolled loop \\
SIMD~\cite[Figure~\ref{code:prefixsum_avx512} of this paper]{hillis1986data} \\
\bottomrule
\end{tabular}

%% file: tables/searching_select.tex
\begin{tabular}{l}
\toprule
loop \\
SIMD \\
\bottomrule
\end{tabular}

%% file: tables/selecting_in_word.tex
\begin{tabular}{l}
\toprule
broadword-sdsl~\cite[Sect. 6.2]{gog2014optimized} \\
broadword-succinct~\cite[Alg. 2]{vigna2008broadword} \\
parallel bits deposit (pdep)~\cite{pandey2017fast} \\
\bottomrule
\end{tabular}

%% file: final_result.tex
\section{Final Result}\label{sec:final_result}

The \code{mutable\_bitmap} class coded in Figure~\ref{code:mbitmap}
(at page~\pageref{code:mbitmap})
can be tuned in many possible ways, depending on which
index and rank/select algorithms are used to maintain its blocks.
For the final examples we are going to illustrate in this section,
we adopt the \emph{best} results from the previous sections.
(Other examples can be obtained by changing these
two components.)
Therefore, our tuning is summarized below.

As index, we use the $b$-ary {\segtree} with SIMD AVX-512
described in Section~\ref{sec:prefix_sums},
as this data structure gives the fastest results for {\Search}
and preserves the runtime of {\Sum}/{\Update} compared to AVX2
(see Figure~\ref{fig:sum_update_search} at page~\pageref{fig:sum_update_search}).
For {\Rank} in a block, we use the algorithm \textsf{builtin+unrolled} for all
bitmap sizes up to $2^{25}$ and then switch to \textsf{builtin+loop}.
We adopt this strategy for both $B=256$ and $B=512$.
For {\Select} in a block, we use \textsf{SIMD+SIMD+pdep} for $B=256$
and \textsf{builtin+loop+pdep} for $B=512$.
Rank and select in a single 64-bit word (i.e., $B=64$)
are respectively done with masking followed by \textsf{builtin} popcount,
and \textsf{pdep}.

\begin{figure}[!t]
\centering
\includegraphics[width=140mm]{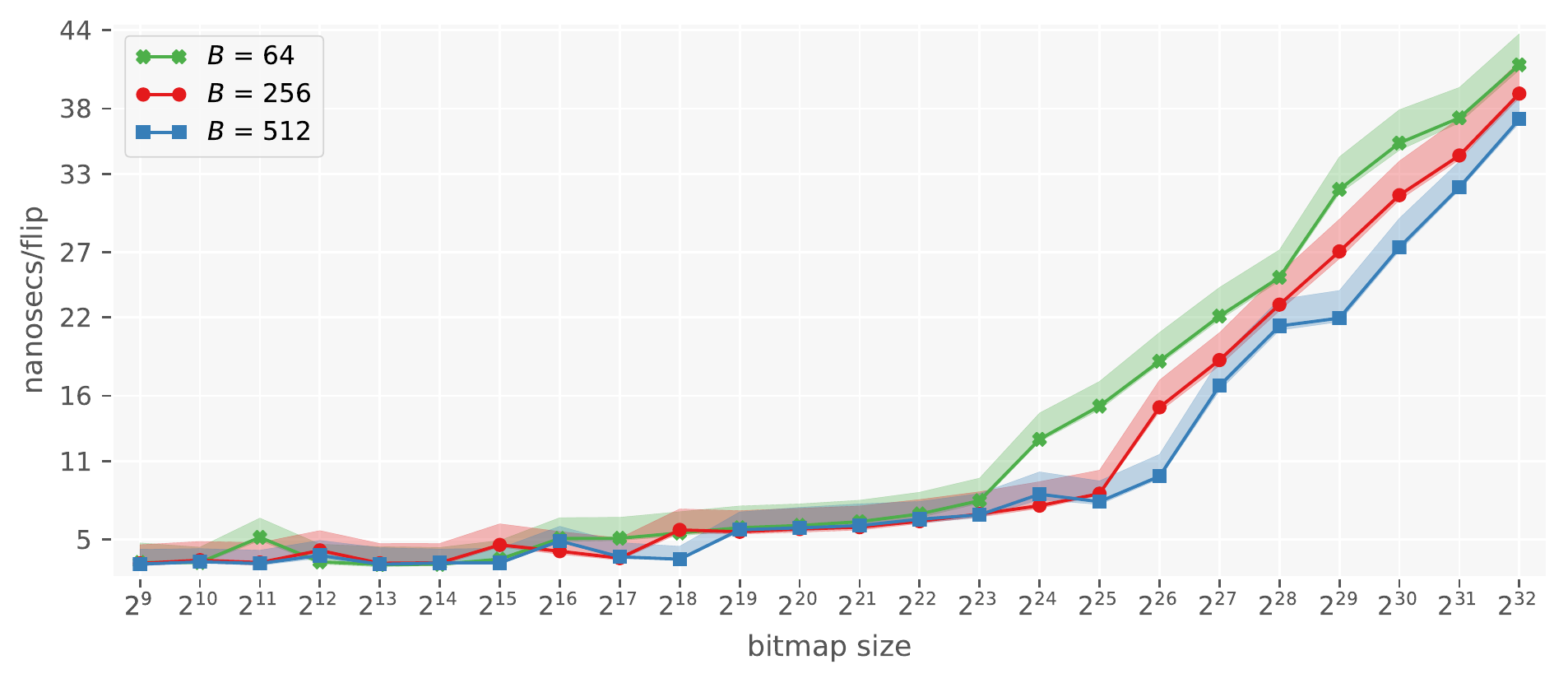}
\includegraphics[width=140mm]{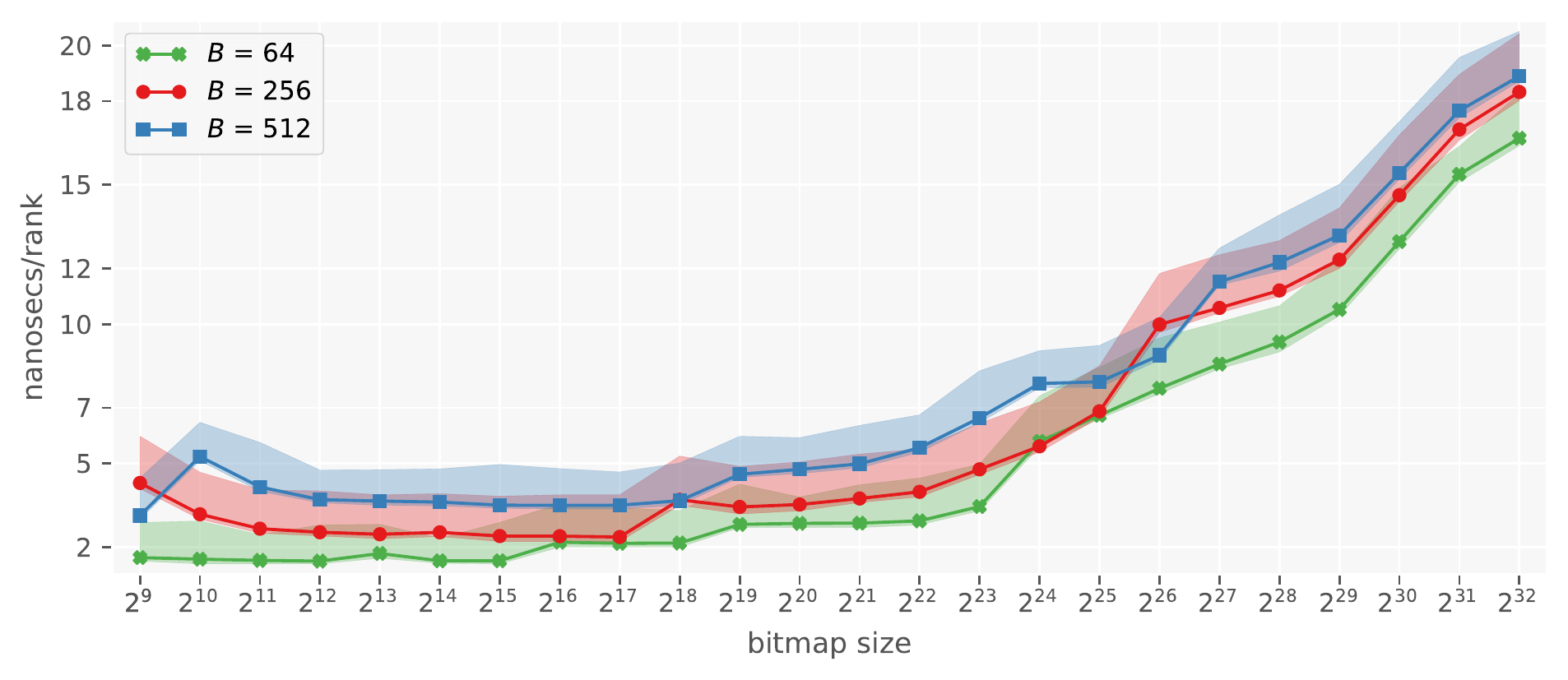}
\includegraphics[width=140mm]{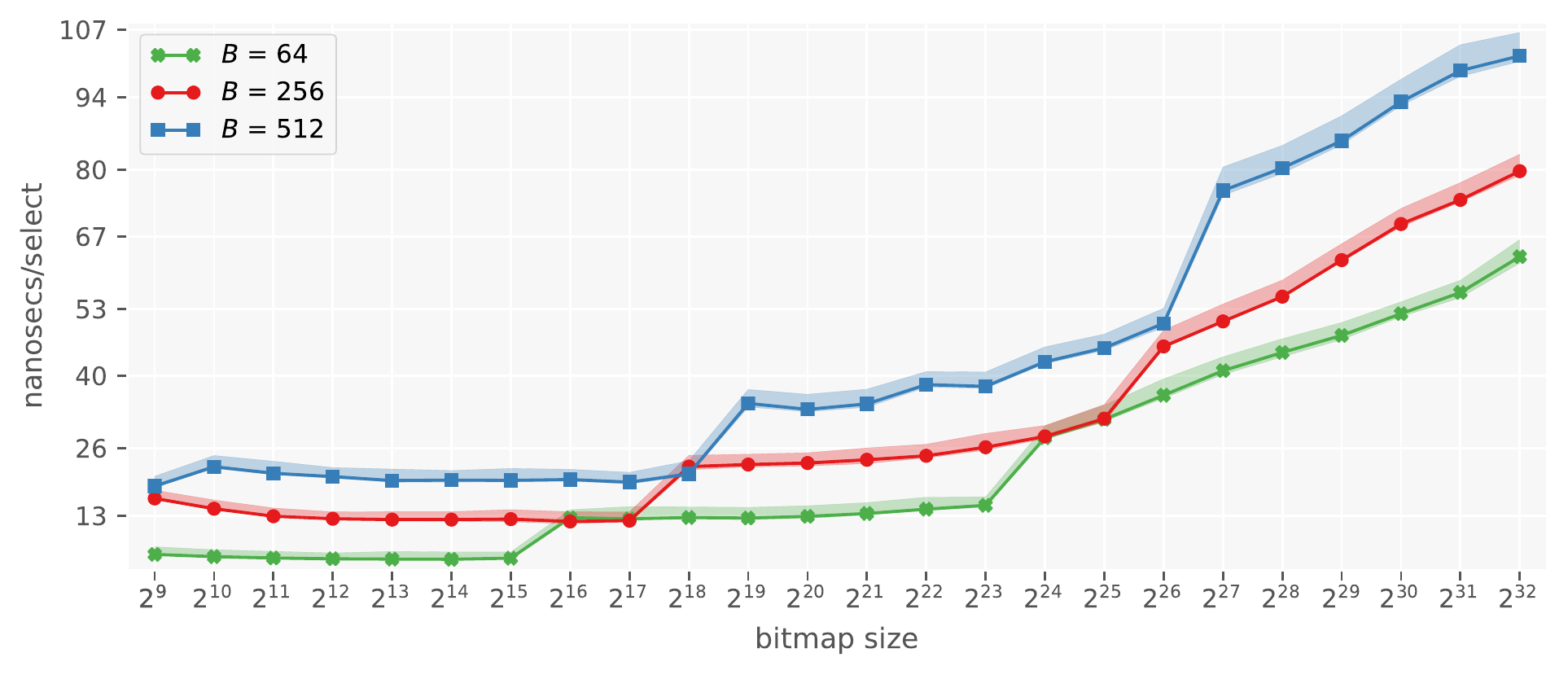}
\caption{Average nanoseconds spent per operation by
the \code{mutable\_bitmap} by varying the block size $B=64, 256, 512$.}
\label{fig:flip_rank_select}
\end{figure}

With the indexing data structure and algorithms fixed,
we show in Figure~\ref{fig:flip_rank_select}
the difference in runtime by varying
the block size $B$.
For all the experiments in this section, we use random
bitmaps of increasing size, whose density is fixed to 0.3,
under a random query workload (as consistently done
in Section~\ref{sec:small_bitmaps}).
The plots for {\Flip} and {\Rank} are very similar for
all block sizes, thus the choice of $B=512$ should be
preferred for the smaller space overhead.
The impact of the block size is evident for
{\Select} instead: increasing the block size
makes the runtime to grow as well for different space/time trade-offs.
In fact, we recall that our solution
with blocks of 256 bits takes 7.2\% extra space,
and just 3.6\% more with blocks of 512 bits.
With the smallest block size, 64,
the extra space becomes 26.7\%.

\begin{figure}[!t]
\centering
\includegraphics[width=140mm]{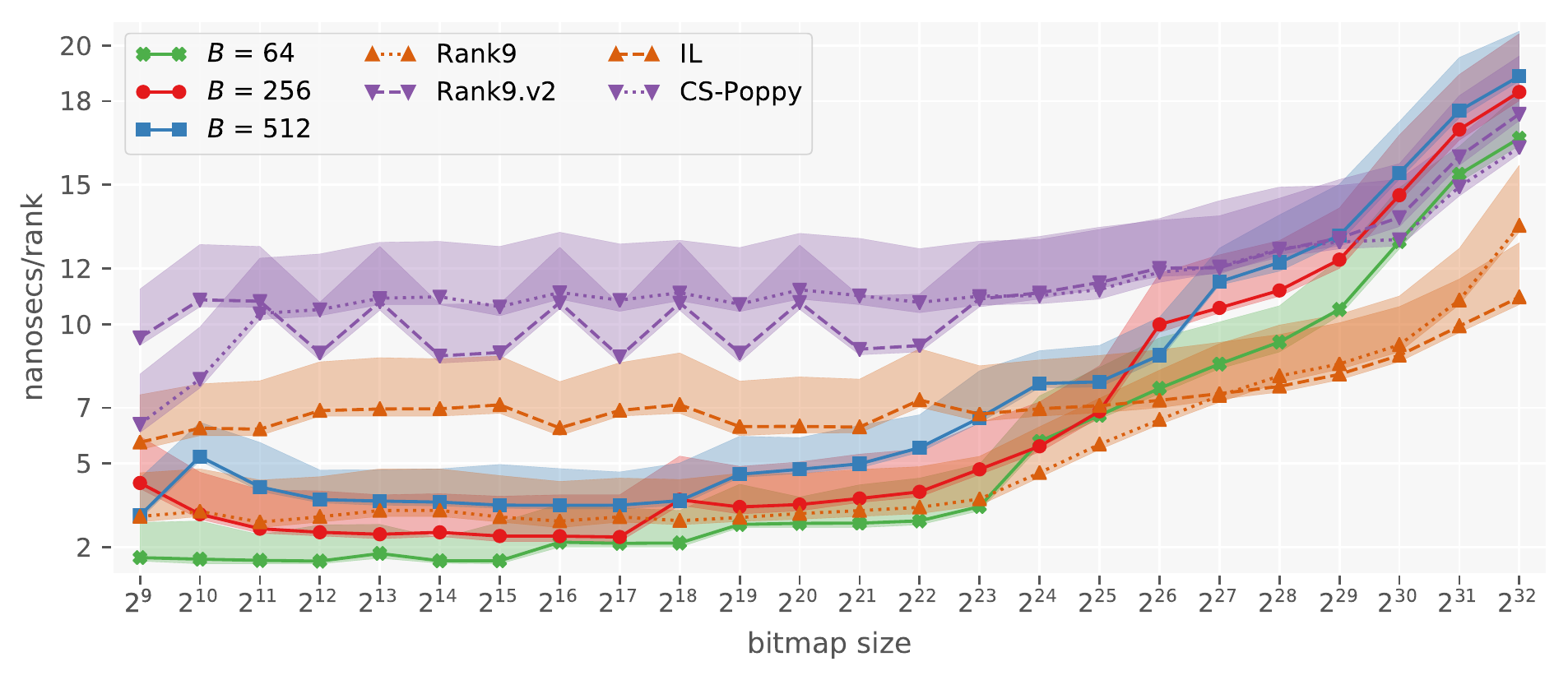}
\includegraphics[width=140mm]{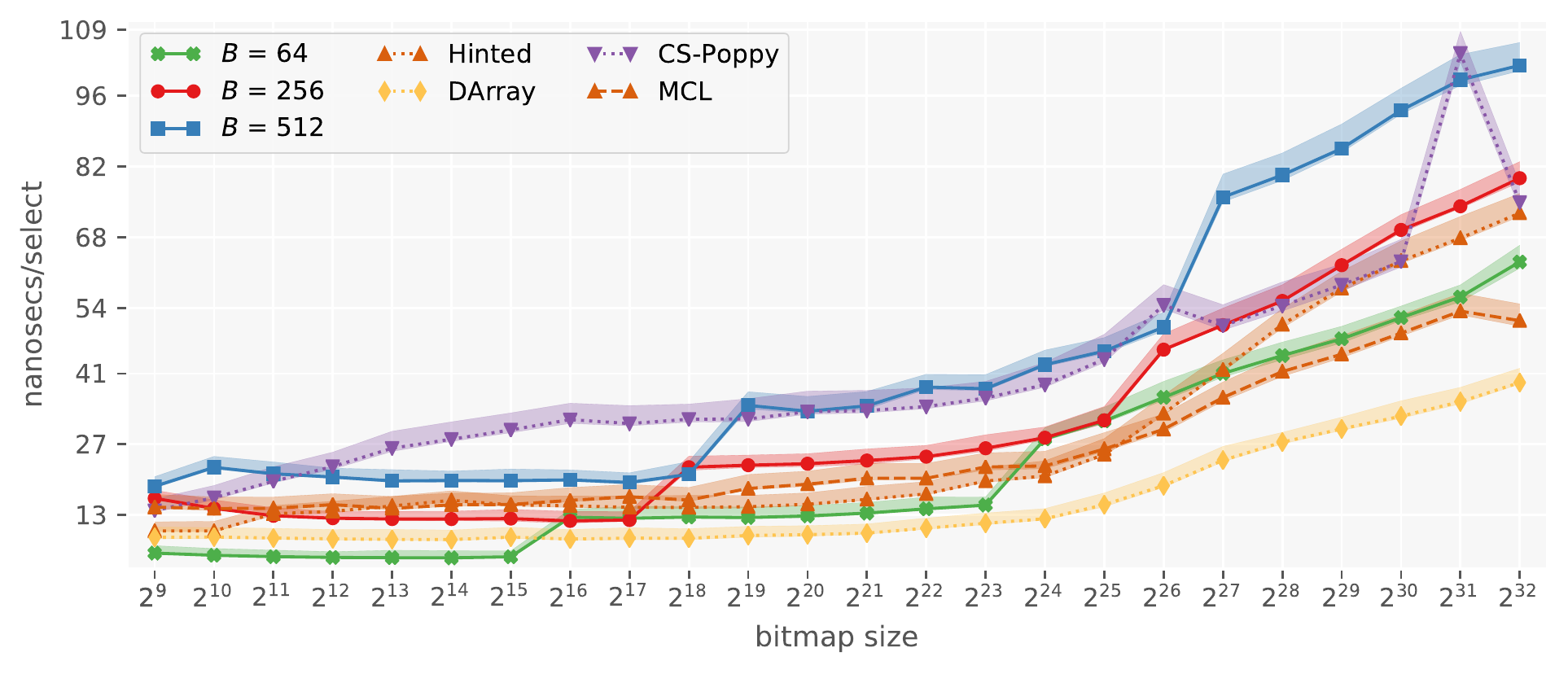}
\caption{Average nanoseconds spent per operation by
the \code{mutable\_bitmap} by varying the block size ($B=64, 256, 512$),
in comparison with other immutable indexes.}
\label{fig:immutable_rank_select}
\end{figure}

\subsection{Comparison against Immutable Indexes}

It is interesting to understand how our mutable index
compares against the \emph{immutable} indexes that we reviewed
in Section~\ref{sec:immutable_bitmaps}, both in the uncompressed and compressed
settings.
This comparison is important to quantify how much the mutability feature
impacts on the performance.

\parag{Uncompressed Indexes}
Figure~\ref{fig:immutable_rank_select} reports the runtime
of some selected uncompressed indexes in comparison
with the mutable bitmap already shown in Figure~\ref{fig:flip_rank_select}.
In order to choose our baselines, we take into account the following
space/time trade-offs:
(1) the most time-efficient solution that requires
a separate index for each query;
(2) the most time-efficient solution that requires a single index
for both queries; and
(3) the most space-efficient solution.
For the first category, we have the combination
\textsf{Rank9+DArray}.
In this case, the extra space results from the sum of the space
of the two indexes,
which is at least 25\% because of \textsf{Rank9}.
We use the implementation available in the \textsf{Succinct}
library. (The implementation of \textsf{Rank9} in \textsf{Sdsl}
performed the same as \textsf{Succinct}'s.)
For the second category, we have the combination
\textsf{Rank9+Hinted} that performs a binary search directly
over the \textsf{Rank9} index to solve {\Select}.
Also in this case we use the \textsf{Succinct}'s implementation
which only requires 3\% extra space on top of \textsf{Rank9},
for a total of 28\% extra.
For the last category, we have \textsf{Poppy}
with \emph{combined sampling} (\textsf{CS-Poppy}),
and we run the implementation by the original authors
which requires 3.4\% extra space to support both queries.

Other baselines available in the \textsf{Sdsl} library
provide additional space/time trade-offs:
\textsf{Rank9.v2} is a more space-efficient version of \textsf{Rank9}
that takes $\approx$6\% extra space;
\textsf{IL} interleaves the original bitmap with rank informations
to help locality of reference, taking 12.5\% extra space;
\textsf{MCL} is a modified Clark's data structure for {\Select},
with low overhead (see Section~\ref{sec:immutable_bitmaps} and Table~\ref{tab:summary},
page~\pageref{tab:summary}).

From Figure~\ref{fig:immutable_rank_select} we see
that the difference in runtime between our mutable
index and the best immutable indexes is not as high as one
would expect.
Thus, our proposal brings
further advantages in that it allows to modify the underlying
bitmap without incurring in a significant runtime penalty and
with even lower space overhead.
More detailed observations are given below.

\begin{itemize}

\item Compared to the combination \textsf{Rank9+DArray},
the mutable index with $B=256$ only introduces a penalty
in runtime for {\Select} on large bitmaps (from $2^{25}$ onward, by $2\times$),
but takes several times less space.
The extra space for \textsf{Rank9} is fixed to 25\% but
the space of \textsf{DArray} depends on
the density of the bitmap because it is a position-based solution.
While we did not observe a meaningful difference in runtime by varying the density,
we measured the space overhead for the densities $[0.1, 0.3, 0.5, 0.7, 0.9]$
and obtained $[5.6\%, 16.9\%, 28.1\%, 39.4\%, 50.6\%]$.
Again, these should be summed to another 25\%.
Note that our mutable index does not depend on the density of the bitmap,
therefore it consumes $4-11\times$ less space.
Lastly, the efficiency gap for {\Select} can be reduced by using a smaller
block size, e.g., $B=64$, with comparable or less space.

\item Compared to the combination \textsf{Rank9+Hinted},
the mutable index with $B=256$ is basically as fast
but also reduces the space overhead by $4\times$, from 28\% to 7\%.

\item Compared to the most space-efficient index, \textsf{CS-Poppy},
the mutable index with $B=512$ is faster (especially for {\Rank})
and takes the same extra space.

\item Concerning the implementations in the \textsf{Sdsl} library,
we see that \textsf{Rank9.v2} performs similarly to \textsf{CS-Poppy}
but consumes more space; \textsf{IL} is $2\times$ more space-efficient than
\textsf{Rank9} but slower for small and medium sized bitmaps;
\textsf{MCL} for {\Select} consumes less space than \textsf{DArray}
but is consistently slower.
It is also possible to binary search the \textsf{IL} layout
to solve {\Select} but the runtime of this approach is $4-5\times$ slower
than that of other approaches for medium and large bitmaps
(therefore, we excluded it from the plots).

\end{itemize}

\parag{Compressed Indexes}
We compare the mutable index with the compressed indexes proposed
by~\citet*{grabowski2018rank}.
(We recall that they aim at compressing the index component as well as the
underlying bitmap. Refer to Section~\ref{sec:immutable_bitmaps},
page~\pageref{par:compressed_bitmaps},
for a description of their compressed layouts.)
For consistency of presentation and methodology,
we test the compressed indexes on random bitmaps
of increasing size with density fixed to 0.3.
As also shown by the original authors
(see \cite[Figure 6]{grabowski2018rank}), the query time and space overhead
of the compressed indexes
is pretty much stable by varying the density of the bitmap
(except possibly, for extremely sparse scenarios, i.e., 0.05 density).
While we cannot expect a great deal of compression
in this setting, our objective here is to
determine the impact of a compressed layout on query time
rather than that of saving space
(indeed, recall that the mutable bitmap is \emph{not} compressed).

\begin{figure}[!t]
\centering
\includegraphics[width=140mm]{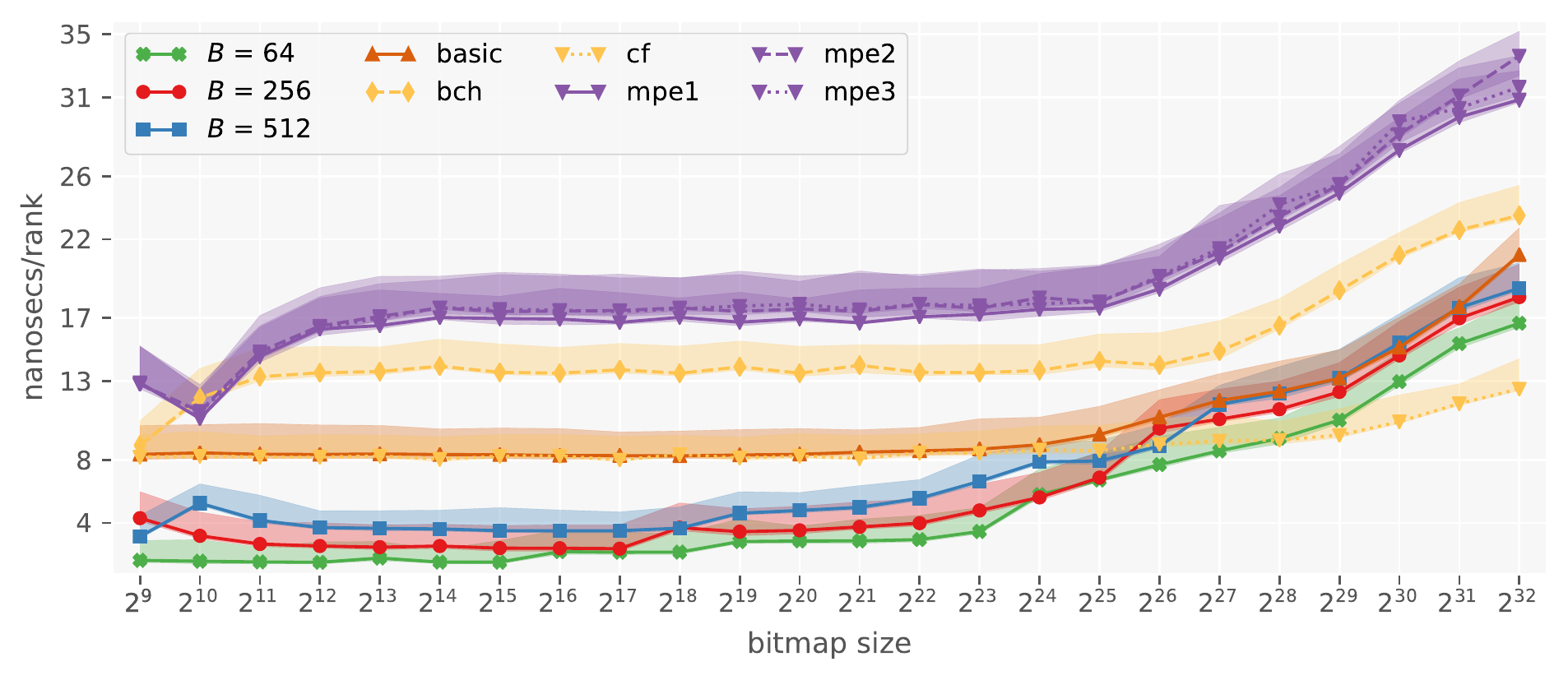}
\includegraphics[width=140mm]{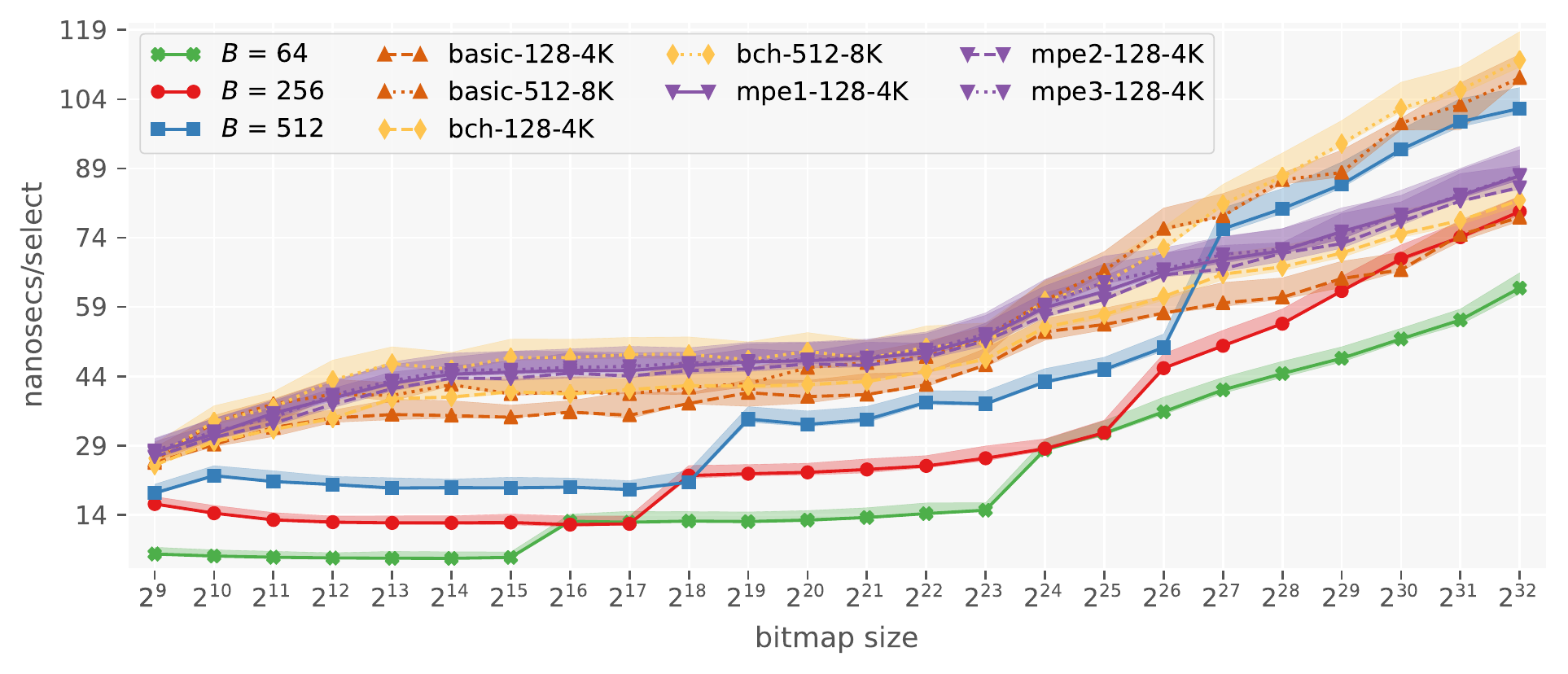}
\caption{Average nanoseconds spent per operation by
the \code{mutable\_bitmap} by varying the block size ($B=64, 256, 512$),
in comparison with other compressed indexes.}
\label{fig:compressed_rank_select}
\end{figure}

Figure~\ref{fig:compressed_rank_select} illustrates the comparison.
We used the same names for the compressed indexes
as given by the authors in their own paper~\cite{grabowski2018rank}.
(From Section~\ref{sec:immutable_bitmaps}, page~\pageref{par:compressed_bitmaps},
we recall that their indexes for {\Select} depend on two parameters,
$\ell$ and $T$, that are also reported in Figure~\ref{fig:compressed_rank_select}.
For example, \textsf{bch-128-4K} indicates the \textsf{bch} layout
with $\ell=128$ and $T=4096$.)
The general trend is that a compressed layout can be faster than
the mutable bitmap for large sizes (larger than, say, $2^{28}$ bits)
because of a better cache exploitation.
A more precise evaluation is as follows.

\begin{itemize}

\item Regarding the {\Rank} query, we determine a space overhead
of: 25\% for \textsf{basic}; 3.7\% for \textsf{bch};
12.5\% for \textsf{cf}; and 4\% for \textsf{mpe1-3}.
Recall that the mutable index has an overhead of 26.7\%, 7.2\%, and 3.6\%,
for $B$ equal to 64, 256, and 512 respectively.
The fastest compressed solution is \textsf{cf}
(``cache friendly''), which is also faster than the mutable index
on bitmap sizes larger than $2^{28}$ bits.
Also, this variant always dominates the
basic variant, being as fast or faster and with less space overhead.
The other solutions, \textsf{bch} and \textsf{mpe1-3},
are consistently slower than the mutable bitmap.

\item Regarding {\Select} instead, we determine a space overhead
of:
31\% for \textsf{basic-128-4K}; 7.6\% for \textsf{basic-512-8K};
20\% for \textsf{bch-128-4K}; 5.1\% for \textsf{bch-512-8K};
20\% for \textsf{mpe-128-4K} (1-3).
In this case, the performance is quite similar for
all the different compressed layouts on bitmaps up to $2^{26}$ bits,
with the \textsf{basic-128-4K} variant being generally faster but with
also the largest space overhead.
On larger bitmaps, the compressed layouts are competitive with
the mutable index for both $B=256$ and $B=512$.

\end{itemize}

As a last note, we remark that the compressed solutions
we take into account here
are optimized for one operation, i.e., the indexes for
{\Rank} and different than those for {\Select}.
This means that, as already
pointed out for other immutable approaches, one has to build both
indexes if both operations are needed for the same bitmap.
The mutable index proposed in this article, instead, supports
three operations (Rank, Select, and Flip) with the same data structure.

%% file: conclusions.tex
\section{Conclusions and Future Work}\label{sec:conclusions}

We have proposed an efficient solution to the problem
of rank/select queries over mutable bitmaps.
The result leverages on the efficiency of two
components:
(1) a searchable prefix-sum data structure, optimized
with SIMD instructions, and
(2) rank/select algorithms over small immutable bitmaps.
Comparison against the best immutable indexes
reveals that a mutable index
can be competitive in query time and consume even less
space.

Future work will focus on making queries even faster,
especially {\Select} that benefited more than {\Rank}
from SIMD optimizations.
Another study could explore the impact of compressing
the blocks of the index for possibly improved space/time
trade-offs in space-constrained applications.

%% file: p.bbl

\begin{thebibliography}{33}


\ifx \showCODEN    \undefined \def \showCODEN     #1{\unskip}     \fi
\ifx \showDOI      \undefined \def \showDOI       #1{#1}\fi
\ifx \showISBNx    \undefined \def \showISBNx     #1{\unskip}     \fi
\ifx \showISBNxiii \undefined \def \showISBNxiii  #1{\unskip}     \fi
\ifx \showISSN     \undefined \def \showISSN      #1{\unskip}     \fi
\ifx \showLCCN     \undefined \def \showLCCN      #1{\unskip}     \fi
\ifx \shownote     \undefined \def \shownote      #1{#1}          \fi
\ifx \showarticletitle \undefined \def \showarticletitle #1{#1}   \fi
\ifx \showURL      \undefined \def \showURL       {\relax}        \fi
\providecommand\bibfield[2]{#2}
\providecommand\bibinfo[2]{#2}
\providecommand\natexlab[1]{#1}
\providecommand\showeprint[2][]{arXiv:#2}

\bibitem[\protect\citeauthoryear{Bentley}{Bentley}{1977}]%
        {bentley1977solutions}
\bibfield{author}{\bibinfo{person}{Jon~Louis Bentley}.}
  \bibinfo{year}{1977}\natexlab{}.
\newblock \showarticletitle{Solutions to Klee's rectangle problems}.
\newblock \bibinfo{journal}{\emph{Unpublished manuscript}}
  (\bibinfo{year}{1977}), \bibinfo{pages}{282--300}.
\newblock


\bibitem[\protect\citeauthoryear{Bentley and Wood}{Bentley and Wood}{1980}]%
        {bentley1980optimal}
\bibfield{author}{\bibinfo{person}{Jon~Louis Bentley} {and}
  \bibinfo{person}{Derick Wood}.} \bibinfo{year}{1980}\natexlab{}.
\newblock \showarticletitle{An optimal worst case algorithm for reporting
  intersections of rectangles}.
\newblock \bibinfo{journal}{\emph{IEEE Trans. Comput.}} \bibinfo{number}{7}
  (\bibinfo{year}{1980}), \bibinfo{pages}{571--577}.
\newblock


\bibitem[\protect\citeauthoryear{Brisaboa, Ladra, and Navarro}{Brisaboa
  et~al\mbox{.}}{2014}]%
        {brisaboa2014compact}
\bibfield{author}{\bibinfo{person}{Nieves~R Brisaboa}, \bibinfo{person}{Susana
  Ladra}, {and} \bibinfo{person}{Gonzalo Navarro}.}
  \bibinfo{year}{2014}\natexlab{}.
\newblock \showarticletitle{Compact representation of web graphs with extended
  functionality}.
\newblock \bibinfo{journal}{\emph{Information Systems}}  \bibinfo{volume}{39}
  (\bibinfo{year}{2014}), \bibinfo{pages}{152--174}.
\newblock


\bibitem[\protect\citeauthoryear{Brisaboa, Luaces, Navarro, and Seco}{Brisaboa
  et~al\mbox{.}}{2013}]%
        {brisaboa2013space}
\bibfield{author}{\bibinfo{person}{Nieves~R Brisaboa},
  \bibinfo{person}{Miguel~R Luaces}, \bibinfo{person}{Gonzalo Navarro}, {and}
  \bibinfo{person}{Diego Seco}.} \bibinfo{year}{2013}\natexlab{}.
\newblock \showarticletitle{Space-efficient representations of rectangle
  datasets supporting orthogonal range querying}.
\newblock \bibinfo{journal}{\emph{Information Systems}} \bibinfo{volume}{38},
  \bibinfo{number}{5} (\bibinfo{year}{2013}), \bibinfo{pages}{635--655}.
\newblock


\bibitem[\protect\citeauthoryear{Clark}{Clark}{1996}]%
        {Clark96}
\bibfield{author}{\bibinfo{person}{David Clark}.}
  \bibinfo{year}{1996}\natexlab{}.
\newblock \emph{\bibinfo{title}{Compact Pat Trees}}.
\newblock \bibinfo{thesistype}{Ph.D. Dissertation}. \bibinfo{school}{University
  of Waterloo}.
\newblock


\bibitem[\protect\citeauthoryear{Corporation}{Corporation}{2020}]%
        {SIMDIntel}
\bibfield{author}{\bibinfo{person}{Intel Corporation}.} \bibinfo{year}{[last
  accessed July 2020]}\natexlab{}.
\newblock \showarticletitle{The Intel Intrinsics Guide,
  \url{https://software.intel.com/sites/landingpage/IntrinsicsGuide}}.
\newblock


\bibitem[\protect\citeauthoryear{Fenwick}{Fenwick}{1994}]%
        {fenwick1994new}
\bibfield{author}{\bibinfo{person}{Peter~M. Fenwick}.}
  \bibinfo{year}{1994}\natexlab{}.
\newblock \showarticletitle{A new data structure for cumulative frequency
  tables}.
\newblock \bibinfo{journal}{\emph{Software: Practice and experience}}
  \bibinfo{volume}{24}, \bibinfo{number}{3} (\bibinfo{year}{1994}),
  \bibinfo{pages}{327--336}.
\newblock


\bibitem[\protect\citeauthoryear{Ferragina, Gonz{\'a}lez, Navarro, and
  Venturini}{Ferragina et~al\mbox{.}}{2009}]%
        {ferragina2009compressed}
\bibfield{author}{\bibinfo{person}{Paolo Ferragina}, \bibinfo{person}{Rodrigo
  Gonz{\'a}lez}, \bibinfo{person}{Gonzalo Navarro}, {and}
  \bibinfo{person}{Rossano Venturini}.} \bibinfo{year}{2009}\natexlab{}.
\newblock \showarticletitle{Compressed text indexes: From theory to practice}.
\newblock \bibinfo{journal}{\emph{Journal of Experimental Algorithmics (JEA)}}
  \bibinfo{volume}{13} (\bibinfo{year}{2009}), \bibinfo{pages}{1--12}.
\newblock


\bibitem[\protect\citeauthoryear{Fredman and Saks}{Fredman and Saks}{1989}]%
        {FS89}
\bibfield{author}{\bibinfo{person}{Michael Fredman} {and}
  \bibinfo{person}{Michael Saks}.} \bibinfo{year}{1989}\natexlab{}.
\newblock \showarticletitle{The cell probe complexity of dynamic data
  structures}. In \bibinfo{booktitle}{\emph{Proceedings of the 21st Annual
  Symposium on Theory of Computing (STOC)}}. \bibinfo{pages}{345--354}.
\newblock


\bibitem[\protect\citeauthoryear{Gog, Beller, Moffat, and Petri}{Gog
  et~al\mbox{.}}{2014}]%
        {gog2014theory}
\bibfield{author}{\bibinfo{person}{Simon Gog}, \bibinfo{person}{Timo Beller},
  \bibinfo{person}{Alistair Moffat}, {and} \bibinfo{person}{Matthias Petri}.}
  \bibinfo{year}{2014}\natexlab{}.
\newblock \showarticletitle{From theory to practice: Plug and play with
  succinct data structures}. In \bibinfo{booktitle}{\emph{Proceedings of the
  13th International Symposium on Experimental Algorithms (SEA)}}. Springer,
  \bibinfo{pages}{326--337}.
\newblock


\bibitem[\protect\citeauthoryear{Gog and Petri}{Gog and Petri}{2014}]%
        {gog2014optimized}
\bibfield{author}{\bibinfo{person}{Simon Gog} {and} \bibinfo{person}{Matthias
  Petri}.} \bibinfo{year}{2014}\natexlab{}.
\newblock \showarticletitle{Optimized succinct data structures for massive
  data}.
\newblock \bibinfo{journal}{\emph{Software: Practice and Experience}}
  \bibinfo{volume}{44}, \bibinfo{number}{11} (\bibinfo{year}{2014}),
  \bibinfo{pages}{1287--1314}.
\newblock


\bibitem[\protect\citeauthoryear{Golynski}{Golynski}{2007}]%
        {golynski2007optimal}
\bibfield{author}{\bibinfo{person}{Alexander Golynski}.}
  \bibinfo{year}{2007}\natexlab{}.
\newblock \showarticletitle{Optimal lower bounds for rank and select indexes}.
\newblock \bibinfo{journal}{\emph{Theoretical Computer Science}}
  \bibinfo{volume}{387}, \bibinfo{number}{3} (\bibinfo{year}{2007}),
  \bibinfo{pages}{348--359}.
\newblock


\bibitem[\protect\citeauthoryear{Gonz{\'a}lez, Grabowski, M{\"a}kinen, and
  Navarro}{Gonz{\'a}lez et~al\mbox{.}}{2005}]%
        {gonzalez2005practical}
\bibfield{author}{\bibinfo{person}{Rodrigo Gonz{\'a}lez},
  \bibinfo{person}{Szymon Grabowski}, \bibinfo{person}{Veli M{\"a}kinen}, {and}
  \bibinfo{person}{Gonzalo Navarro}.} \bibinfo{year}{2005}\natexlab{}.
\newblock \showarticletitle{Practical implementation of rank and select
  queries}. In \bibinfo{booktitle}{\emph{Poster Proceedings of the 4th Workshop
  on Experimental and Efficient Algorithms (WEA)}}. \bibinfo{pages}{27--38}.
\newblock


\bibitem[\protect\citeauthoryear{Grabowski and Raniszewski}{Grabowski and
  Raniszewski}{2018}]%
        {grabowski2018rank}
\bibfield{author}{\bibinfo{person}{Szymon Grabowski} {and}
  \bibinfo{person}{Marcin Raniszewski}.} \bibinfo{year}{2018}\natexlab{}.
\newblock \showarticletitle{Rank and select: Another lesson learned}.
\newblock \bibinfo{journal}{\emph{Information Systems}}  \bibinfo{volume}{73}
  (\bibinfo{year}{2018}), \bibinfo{pages}{25--34}.
\newblock


\bibitem[\protect\citeauthoryear{Grossi and Ottaviano}{Grossi and
  Ottaviano}{2013}]%
        {grossi2013design}
\bibfield{author}{\bibinfo{person}{Roberto Grossi} {and}
  \bibinfo{person}{Giuseppe Ottaviano}.} \bibinfo{year}{2013}\natexlab{}.
\newblock \showarticletitle{Design of practical succinct data structures for
  large data collections}. In \bibinfo{booktitle}{\emph{Proceedings of the 12th
  International Symposium on Experimental Algorithms (SEA)}}. Springer,
  \bibinfo{pages}{5--17}.
\newblock


\bibitem[\protect\citeauthoryear{Hillis and Steele}{Hillis and Steele}{1986}]%
        {hillis1986data}
\bibfield{author}{\bibinfo{person}{W.~Daniel Hillis} {and}
  \bibinfo{person}{Guy~L. Steele}.} \bibinfo{year}{1986}\natexlab{}.
\newblock \showarticletitle{Data parallel algorithms}.
\newblock \bibinfo{journal}{\emph{Commun. ACM}} \bibinfo{volume}{29},
  \bibinfo{number}{12} (\bibinfo{date}{Dec.} \bibinfo{year}{1986}),
  \bibinfo{pages}{1170–1183}.
\newblock


\bibitem[\protect\citeauthoryear{Jacobson}{Jacobson}{1988}]%
        {jacobson1988succinct}
\bibfield{author}{\bibinfo{person}{Guy~Joseph Jacobson}.}
  \bibinfo{year}{1988}\natexlab{}.
\newblock \emph{\bibinfo{title}{Succinct static data structures}}.
\newblock \bibinfo{thesistype}{Ph.D. Dissertation}. \bibinfo{school}{Carnegie
  Mellon University}.
\newblock


\bibitem[\protect\citeauthoryear{Kanda, Morita, and Fuketa}{Kanda
  et~al\mbox{.}}{2017}]%
        {kanda2017string}
\bibfield{author}{\bibinfo{person}{Shunsuke Kanda}, \bibinfo{person}{Kazuhiro
  Morita}, {and} \bibinfo{person}{Masao Fuketa}.}
  \bibinfo{year}{2017}\natexlab{}.
\newblock \showarticletitle{Practical string dictionary compression using
  string dictionary encoding}. In \bibinfo{booktitle}{\emph{Proceedings of the
  3rd International Conference on Big Data Innovations and Applications
  (Innovate-Data)}}. \bibinfo{pages}{1--8}.
\newblock


\bibitem[\protect\citeauthoryear{Marchini and Vigna}{Marchini and
  Vigna}{2020}]%
        {marchini2020compact}
\bibfield{author}{\bibinfo{person}{Stefano Marchini} {and}
  \bibinfo{person}{Sebastiano Vigna}.} \bibinfo{year}{2020}\natexlab{}.
\newblock \showarticletitle{Compact Fenwick trees for dynamic ranking and
  selection}.
\newblock \bibinfo{journal}{\emph{Software: Practice and Experience}}
  \bibinfo{volume}{50}, \bibinfo{number}{7} (\bibinfo{year}{2020}),
  \bibinfo{pages}{1184--1202}.
\newblock


\bibitem[\protect\citeauthoryear{Mart{\'\i}nez-Prieto, Brisaboa, C{\'a}novas,
  Claude, and Navarro}{Mart{\'\i}nez-Prieto et~al\mbox{.}}{2016}]%
        {martinez2016practical}
\bibfield{author}{\bibinfo{person}{Miguel~A Mart{\'\i}nez-Prieto},
  \bibinfo{person}{Nieves Brisaboa}, \bibinfo{person}{Rodrigo C{\'a}novas},
  \bibinfo{person}{Francisco Claude}, {and} \bibinfo{person}{Gonzalo Navarro}.}
  \bibinfo{year}{2016}\natexlab{}.
\newblock \showarticletitle{Practical compressed string dictionaries}.
\newblock \bibinfo{journal}{\emph{Information Systems}}  \bibinfo{volume}{56}
  (\bibinfo{year}{2016}), \bibinfo{pages}{73--108}.
\newblock


\bibitem[\protect\citeauthoryear{Miltersen}{Miltersen}{2005}]%
        {miltersen2005lower}
\bibfield{author}{\bibinfo{person}{Peter~Bro Miltersen}.}
  \bibinfo{year}{2005}\natexlab{}.
\newblock \showarticletitle{Lower bounds on the size of selection and rank
  indexes}. In \bibinfo{booktitle}{\emph{Proceedings of the 16th Annual
  ACM-SIAM Symposium on Discrete Algorithms (SODA)}},
  Vol.~\bibinfo{volume}{23}. Citeseer, \bibinfo{pages}{11--12}.
\newblock


\bibitem[\protect\citeauthoryear{Mu{\l}a, Kurz, and Lemire}{Mu{\l}a
  et~al\mbox{.}}{2017}]%
        {mula2017faster}
\bibfield{author}{\bibinfo{person}{Wojciech Mu{\l}a}, \bibinfo{person}{Nathan
  Kurz}, {and} \bibinfo{person}{Daniel Lemire}.}
  \bibinfo{year}{2017}\natexlab{}.
\newblock \showarticletitle{Faster population counts using AVX2 instructions}.
\newblock \bibinfo{journal}{\emph{Comput. J.}} \bibinfo{volume}{61},
  \bibinfo{number}{1} (\bibinfo{year}{2017}), \bibinfo{pages}{111--120}.
\newblock


\bibitem[\protect\citeauthoryear{Navarro and Providel}{Navarro and
  Providel}{2012}]%
        {navarro2012fast}
\bibfield{author}{\bibinfo{person}{Gonzalo Navarro} {and}
  \bibinfo{person}{Eliana Providel}.} \bibinfo{year}{2012}\natexlab{}.
\newblock \showarticletitle{Fast, small, simple rank/select on bitmaps}. In
  \bibinfo{booktitle}{\emph{Proceedings of the 11th International Symposium on
  Experimental Algorithms (SEA)}}. Springer, \bibinfo{pages}{295--306}.
\newblock


\bibitem[\protect\citeauthoryear{Okanohara and Sadakane}{Okanohara and
  Sadakane}{2007}]%
        {okanohara2007practical}
\bibfield{author}{\bibinfo{person}{Daisuke Okanohara} {and}
  \bibinfo{person}{Kunihiko Sadakane}.} \bibinfo{year}{2007}\natexlab{}.
\newblock \showarticletitle{Practical entropy-compressed rank/select
  dictionary}. In \bibinfo{booktitle}{\emph{Proceedings of the 9th Workshop on
  Algorithm Engineering and Experiments (ALENEX)}}. SIAM,
  \bibinfo{pages}{60--70}.
\newblock


\bibitem[\protect\citeauthoryear{Pandey, Bender, and Johnson}{Pandey
  et~al\mbox{.}}{2017}]%
        {pandey2017fast}
\bibfield{author}{\bibinfo{person}{Prashant Pandey}, \bibinfo{person}{Michael~A
  Bender}, {and} \bibinfo{person}{Rob Johnson}.}
  \bibinfo{year}{2017}\natexlab{}.
\newblock \showarticletitle{A fast x86 implementation of select}.
\newblock \bibinfo{journal}{\emph{arXiv preprint arXiv:1706.00990}}
  (\bibinfo{year}{2017}).
\newblock


\bibitem[\protect\citeauthoryear{Pibiri and Venturini}{Pibiri and
  Venturini}{2020a}]%
        {pibiri2020practical}
\bibfield{author}{\bibinfo{person}{Giulio~Ermanno Pibiri} {and}
  \bibinfo{person}{Rossano Venturini}.} \bibinfo{year}{2020}\natexlab{a}.
\newblock \showarticletitle{Practical trade-offs for the prefix-sum problem}.
\newblock \bibinfo{journal}{\emph{Software: Practice and Experience}}
  \bibinfo{volume}{To Appear.} (\bibinfo{year}{2020}).
\newblock


\bibitem[\protect\citeauthoryear{Pibiri and Venturini}{Pibiri and
  Venturini}{2020b}]%
        {pibiri2020techniques}
\bibfield{author}{\bibinfo{person}{Giulio~Ermanno Pibiri} {and}
  \bibinfo{person}{Rossano Venturini}.} \bibinfo{year}{2020}\natexlab{b}.
\newblock \showarticletitle{Techniques for Inverted Index Compression}.
\newblock \bibinfo{journal}{\emph{ACM Computing Surveys (CSUR)}}
  \bibinfo{volume}{53}, \bibinfo{number}{6}, Article \bibinfo{articleno}{125}
  (\bibinfo{year}{2020}), \bibinfo{numpages}{36}~pages.
\newblock


\bibitem[\protect\citeauthoryear{Prezza}{Prezza}{2017}]%
        {prezza2017framework}
\bibfield{author}{\bibinfo{person}{Nicola Prezza}.}
  \bibinfo{year}{2017}\natexlab{}.
\newblock \showarticletitle{A framework of dynamic data structures for string
  processing}. In \bibinfo{booktitle}{\emph{Proceedings of the 16th
  International Symposium on Experimental Algorithms (SEA)}},
  Vol.~\bibinfo{volume}{75}. \bibinfo{pages}{11:1--11:15}.
\newblock


\bibitem[\protect\citeauthoryear{Raman, Raman, and Satti}{Raman
  et~al\mbox{.}}{2007}]%
        {raman2007succinct}
\bibfield{author}{\bibinfo{person}{Rajeev Raman}, \bibinfo{person}{Venkatesh
  Raman}, {and} \bibinfo{person}{Srinivasa~Rao Satti}.}
  \bibinfo{year}{2007}\natexlab{}.
\newblock \showarticletitle{Succinct indexable dictionaries with applications
  to encoding k-ary trees, prefix sums and multisets}.
\newblock \bibinfo{journal}{\emph{ACM Transactions on Algorithms (TALG)}}
  \bibinfo{volume}{3}, \bibinfo{number}{4} (\bibinfo{year}{2007}),
  \bibinfo{pages}{43--es}.
\newblock


\bibitem[\protect\citeauthoryear{Vigna}{Vigna}{2008}]%
        {vigna2008broadword}
\bibfield{author}{\bibinfo{person}{Sebastiano Vigna}.}
  \bibinfo{year}{2008}\natexlab{}.
\newblock \showarticletitle{Broadword implementation of rank/select queries}.
  In \bibinfo{booktitle}{\emph{Proceedings of the 7th International Workshop on
  Experimental and Efficient Algorithms (WEA)}}. Springer,
  \bibinfo{pages}{154--168}.
\newblock


\bibitem[\protect\citeauthoryear{Vigna}{Vigna}{2020}]%
        {sux}
\bibfield{author}{\bibinfo{person}{Sebastiano Vigna}.} \bibinfo{year}{[last
  accessed July 2020]}\natexlab{}.
\newblock \showarticletitle{The \textsf{Sux} library,
  \url{https://github.com/vigna/sux}}.
\newblock


\bibitem[\protect\citeauthoryear{Zhang, Lim, Leis, Andersen, Kaminsky, Keeton,
  and Pavlo}{Zhang et~al\mbox{.}}{2018}]%
        {zhang2018surf}
\bibfield{author}{\bibinfo{person}{Huanchen Zhang}, \bibinfo{person}{Hyeontaek
  Lim}, \bibinfo{person}{Viktor Leis}, \bibinfo{person}{David~G Andersen},
  \bibinfo{person}{Michael Kaminsky}, \bibinfo{person}{Kimberly Keeton}, {and}
  \bibinfo{person}{Andrew Pavlo}.} \bibinfo{year}{2018}\natexlab{}.
\newblock \showarticletitle{{SuRF}: Practical range query filtering with fast
  succinct tries}. In \bibinfo{booktitle}{\emph{Proceedings of the 2018
  International Conference on Management of Data (SIGMOD)}}.
  \bibinfo{pages}{323--336}.
\newblock


\bibitem[\protect\citeauthoryear{Zhou, Andersen, and Kaminsky}{Zhou
  et~al\mbox{.}}{2013}]%
        {zhou2013space}
\bibfield{author}{\bibinfo{person}{Dong Zhou}, \bibinfo{person}{David~G
  Andersen}, {and} \bibinfo{person}{Michael Kaminsky}.}
  \bibinfo{year}{2013}\natexlab{}.
\newblock \showarticletitle{Space-efficient, high-performance rank and select
  structures on uncompressed bit sequences}. In
  \bibinfo{booktitle}{\emph{Proceedings of the 12th International Symposium on
  Experimental Algorithms (SEA)}}. Springer, \bibinfo{pages}{151--163}.
\newblock


\end{thebibliography}
